\documentclass[runningheads]{llncs}

\newif\ifFullVersion 
\FullVersiontrue
\newif\ifHighlightVersionDiff
\newcommand{\myfootnotesize}{\fontsize{8pt}{10pt}\selectfont}
\usepackage[inline]{enumitem}

\newcommand{\tinier}{\fontsize{6pt}{7.5pt}\selectfont}

\usepackage[nocompress]{cite}

\usepackage{hyperref}

\usepackage[T1]{fontenc}
\usepackage[utf8]{inputenc}

\usepackage{amssymb}

\usepackage{wrapfig}
\usepackage{microtype}
\usepackage{xspace}
\usepackage{mathtools}
\usepackage{calc}
\usepackage{array}
\usepackage{caption}
\usepackage{subcaption}
\usepackage{booktabs}   %
\usepackage{polytable}

\usepackage{comment}
\usepackage{tikz}
\usepackage{tikz-cd}
\usetikzlibrary{positioning,fit,calc}

\makeatletter
\def\qedmath{%
\relax\ifmmode
   \@badmath
\else
   \bgroup
   \setlength{\topsep}{0pt}%
   \setlength{\parsep}{0pt}%
   \begin{trivlist}%
      \@beginparpenalty\predisplaypenalty
      \@endparpenalty\postdisplaypenalty
      \item[]\leavevmode
      \hbox to\linewidth\bgroup\relax\hfil$\displaystyle
      \bgroup
\fi
}
\def\endqedmath{%
   \relax\ifmmode
   \egroup $\hfil \egroup\\[-\baselineskip]
   \hbox{}\hfill\llap{\phantom{0}\qed}
      \end{trivlist}\egroup
   \else
      \@badmath
   \fi
}
\def\qedflmath{\relax \ifmmode \@badmath
   \else\begin{trivlist}%
         \@beginparpenalty\predisplaypenalty
         \@endparpenalty\postdisplaypenalty
         \item[]\leavevmode \hb@xt@\linewidth\bgroup $\m@th\displaystyle %
            \hskip\mathindent\bgroup
            \fi}
            \def\endqedflmath{\relax\ifmmode
            \egroup $%
         \egroup \\[-\baselineskip]\hbox{}\hfill\llap{\phantom{0}\qed}
      \end{trivlist}%
   \else \@badmath
   \fi}
\makeatother

\usepackage{xparse}
\ExplSyntaxOn
\NewExpandableDocumentCommand{\caseswitch}{O{}mm}
 {
  \str_case_e:nnF { #2 } { #3 } { #1 }
 }
\ExplSyntaxOff

\definecolor{darkpastelgreen}{rgb}{0.01, 0.75, 0.24}
\definecolor{seagreen}{rgb}{0.18, 0.55, 0.34}
\definecolor{richlavender}{rgb}{0.67, 0.38, 0.8}
\definecolor{darkcyan}{rgb}{0.0, 0.55, 0.55}
\definecolor{jade}{rgb}{0.0, 0.66, 0.42}
\definecolor{caribbeangreen}{rgb}{0.0, 0.8, 0.6}

\definecolor{dkgreen}{HTML}{006329}

\newcommand{\mi}[1]{\ensuremath{\mathit{#1}}}

\newcommand{\ms}[1]{\ensuremath{\mathsf{#1}}}

\newcommand{\foreignWord}[1]{#1}
\newcommand{\eg}{\foreignWord{e.g.}}
\newcommand{\ie}{\foreignWord{i.e.}}

\newcommand{\etal}{\foreignWord{et al.}}

\newcommand{\Gs}{\sigma}

\newcommand{\GTH}{\Theta}

\newcommand{\GO}{\Omega}

\newcommand{\x}{\times}

\newenvironment{tarray}[2][c]{%
\settowidth{\dimen1}{${} = {}$}%
   \setlength{\arraycolsep}{0.125\dimen1}
   \hspace{-0.125\dimen1}\array[#1]{#2}\relax
}
{%
   \endarray\hspace{-0.125\dimen1}
}
\newcommand{\bb}{\tarray{lllll}}

\newcommand{\bbt}{\tarray[t]{lllll}}

\newcommand{\ee}{\endtarray}

\newcommand{\pto}{\rightharpoonup}

\newcommand{\key}[1]{\ensuremath{\mathbf{#1}}} %

\newcommand{\LET}{\key{let}}
\newcommand{\IN}{\key{in}}

\newcommand{\dontcare}{%
   \settowidth{\dimen2}{v}%
   \hspace{0.2\dimen2}\rule{0.9\dimen2}{0.5pt}\hspace{0.2\dimen2}}
\newcommand{\A}{\;}

\newcommand{\shortequals}{%
   {\rlap{\rule[0.14em]{.25em}{0.5pt}}{\rule[0.34em]{.25em}{0.5pt}}}
}

\newcommand{\doubleequals}{\ensuremath{\mathrel{%
         \shortequals\,\shortequals}}}
\newcommand{\ndoubleequals}{%
   \ensuremath{\not\expandafter\doubleequals}}

\makeatletter
\newcounter{pe@nocounter}
\newcommand{\peano}[1]{%
   \setcounter{pe@nocounter}{#1}
   \pe@no}
\newcommand{\pe@no}{%
   \ifnum\value{pe@nocounter}>1
      \addtocounter{pe@nocounter}{-1}%
      \con{S} \A (\expandafter\pe@no)%
   \else%
      \ifnum\value{pe@nocounter}>0%
         \con{S} \A \con{Z}%
      \else%
         \con{Z}%
      \fi%
   \fi%
}

\newcommand{\con}[1]{\ms{#1}}
\newcommand{\var}[1]{\mi{#1}}

\newcommand{\TT}{\con{True}}
\newcommand{\FF}{\con{False}}

\newcommand{\str}[1]{\texttt{"#1"}}

\DeclareMathOperator{\dom}{dom}

\definecolor{bxcolor}{rgb}{0.05,0.15,0.76}

\makeatletter
\newif\ifbxgrayed@ut
\newcommand{\grayout}{\bxgrayed@uttrue\color{gray}}
\newcommand{\grayoutend}{\bxgrayed@utfalse}

\newcommand{\bxcolor}{\ifbxgrayed@ut\color{gray}\else\color{bxcolor}\fi}
\newcommand{\textbxcolor}[1]{\ifbxgrayed@ut\textcolor{gray}{#1}\else\textcolor{bxcolor}{#1}\fi}

\makeatother

\newcommand{\infer}[2]{\frac{\displaystyle#2}{\displaystyle#1}}

\newcommand{\GET}{\ensuremath{\mathit{get}}\xspace}
\newcommand{\PUT}{\ensuremath{\mathit{put}}\xspace}

\newcommand{\UnTagS}{\ensuremath{\var{unTagS}}}
\newcommand{\UnTag}{\ensuremath{\var{unTag}}}

\usepackage{stmaryrd}
\newcommand{\sem}[1]{\llbracket#1\rrbracket}

\newcommand{\UnitType}{\ensuremath{\mathbf{1}}}
\newcommand{\InL}{\ensuremath{\con{InL}}}
\newcommand{\InR}{\ensuremath{\con{InR}}}

\usepackage{thmtools}

\spnewtheorem*{myremark}{Remark}{\itshape}{\rmfamily}

\makeatletter
\@spothm{nonexample}[example]{Nonexample}{\itshape}{\rmfamily}
\makeatother

\makeatletter
\newenvironment{myproof}{%
  \par
  \topsep 7\p@ \@plus2\p@ \@minus4\p@
  \addvspace{\topsep}%
  \noindent\textit{Proof.}\hskip\labelsep
  \ignorespaces
}{%
  \unskip\nobreak\hfill$\qed$%
  \par
  \addvspace{\topsep}%
}
\makeatother

\usepackage{cleveref}

\crefname{lemma}{Lemma}{Lemmas}
\crefname{theorem}{Theorem}{Theorems}
\crefname{section}{Sect.}{Sects.}
\crefname{example}{Example}{Examples}
\crefname{figure}{Fig.}{Figs.}

\crefname{appendix}{Appendix}{Appendices}

\usepackage{xcolor}

\usepackage[misc]{ifsym}

\definecolor{diffcolor}{rgb}{0.18, 0.55, 0.34}

\ifFullVersion
  \ifHighlightVersionDiff
    \NewDocumentCommand\fullversion{m}{\textcolor{diffcolor}{#1}\ignorespaces}%
    \specialcomment{fullversionblock}{\begingroup\color{diffcolor}}{\endgroup}%
  \else 
    \NewDocumentCommand\fullversion{m}{#1\ignorespaces}%
    \includecomment{fullversionblock}%
  \fi 
  \NewDocumentCommand\confversion{m}{\ignorespaces}%
  \excludecomment{confversionblock}%
\else 
  \ifHighlightVersionDiff 
    \NewDocumentCommand\confversion{m}{\textcolor{diffcolor}{#1}\ignorespaces}%
    \specialcomment{confversionblock}{\begingroup\color{diffcolor}}{\endgroup}%
  \else 
    \NewDocumentCommand\confversion{m}{#1\ignorespaces}%
    \includecomment{confversionblock}%
  \fi 
  \NewDocumentCommand\fullversion{m}{\ignorespaces}%
  \excludecomment{fullversionblock}%
\fi

\makeatletter

\@ifundefined{mathindent}%
  {\newdimen\mathindent\mathindent\leftmargini}%
  {}%

\makeatother
\newlength{\blanklineskip}
\DeclareMathOperator{\Disjoint}{Disj}

\newcommand{\Carrier}[1]{|#1|}

\newcommand{\pstate}{\pstateHead state\xspace}
\newcommand{\pstates}{\pstateHead states\xspace}

\newcommand{\Pstates}{\PstateHead states\xspace}
\newcommand{\fstate}{\fstateHead state\xspace}
\newcommand{\fstates}{\fstateHead states\xspace}

\newcommand{\pstateHead}{partially-specified\xspace}
\newcommand{\fstateHead}{proper\xspace}
\newcommand{\PstateHead}{Partially-specified\xspace}

\newcommand{\spstate}{strictly \pstate}
\newcommand{\spstates}{strictly \pstates}

\newcommand{\Spstates}{Strictly \pstates}

\newcommand{\our}{partial-state\xspace}
\newcommand{\Our}{Partial-state\xspace}
\newcommand{\OUR}{Partial-State\xspace}
\newcommand{\ourshort}{ps}

\newcommand{\IPoset}{i-poset\xspace}
\newcommand{\IPosets}{i-posets\xspace}
\newcommand{\IDonName}{\mathit{I}}
\newcommand{\IDonSet}[1]{\mathit{I}_{#1}}
\newcommand{\IDon}[2]{#1 \in \IDonSet{#2}}

\newcommand{\GenIPoset}[2]{\ensuremath{\mathsf{G}_{#1,#2}}}

\newcommand{\initiator}{ps-initiator\xspace}
\newcommand{\INITIATOR}{Ps-Initiator\xspace}
\newcommand{\initiators}{ps-initiators\xspace}
\newcommand{\INITIATORs}{Ps-Initiators\xspace}
\newcommand{\TaskInit}{\ensuremath{\var{init}_\mathrm{tasks}}\xspace}
\newcommand{\ApplyUpd}[2]{\ensuremath{#1 \cdot #2}}
\newcommand{\ApplyUpdD}[2]{\ensuremath{\sem{#1}(#2)}}
\newcommand{\SemUpd}[1]{\ensuremath{\sem{#1}}}

\newcommand{\Ran}[2]{\ensuremath{\mathrm{ran}_{#1}(#2)}}

\NewDocumentCommand\DUP{}{\var{dup}}
\NewDocumentCommand\VName{m}{\textsc{#1}}

\NewDocumentCommand\LensName{}{\con{Lens}}
\NewDocumentCommand\Lens{m m}{\LensName\;#1\;#2}
\NewDocumentCommand\MALensName{}{\con{Lens}^{\le}}
\NewDocumentCommand\MALens{m m}{\MALensName\;#1\;#2}

\NewDocumentCommand\putarrow{O{""} O{10pt}}{%
\arrow[from=r, mapsto, #1]%
\arrow[from=r, phantom, ""{coordinate, name=AppBpp}]%
\arrow[from=u, mapsto, rounded corners=#2, to path={(\tikztostart.east) -| (AppBpp)\tikztonodes -- (\tikztotarget)}]%
}

\usetikzlibrary{backgrounds}
\tikzstyle{none}=[inner sep=0mm]
\usetikzlibrary{arrows}
\usetikzlibrary{shapes,shapes.geometric,shapes.misc}
\tikzstyle{box}=[fill=white, draw=black, shape=rectangle, minimum width=0.75cm, minimum height=1cm]
\begin{document}
\setlength{\jot}{0pt}

\title{Lenses for Partially-Specified States\ifFullVersion\\(Extended Version)\fi}

\author{Kazutaka Matsuda\inst{1}\orcidID{0000-0002-9747-4899}\textsuperscript{(\Letter)} \and Minh Nguyen \inst{2}\orcidID{0000-0003-3845-9928} \and Meng Wang \inst{2}\orcidID{0000-0001-7780-630X}}
\institute{Tohoku University,
Sendai 980-8579, Japan.\\
\email{kztk@tohoku.ac.jp}
\and University of Bristol, Bristol BS8 1TH, United Kingdom.
}

\authorrunning{K. Matsuda et al.}

\maketitle

\begin{abstract}

A bidirectional transformation is a pair of transformations satisfying certain well-behavedness properties: one maps source data into view data, and the other translates changes on the view back to the source. However, when multiple views share a source, an update on one view may affect the others, making it hard to maintain correspondence while preserving the user's update, especially when multiple views are changed at once. Ensuring these properties within a compositional framework is even more challenging.
In this paper, we propose \emph{\our{} lenses}, which allow source and view states to be partially specified to precisely represent the user's update intentions. 
These intentions are partially ordered, providing clear semantics for merging intentions of updates coming from multiple views and a refined notion of update preservation compatible with this merging. 
We formalize \our{} lenses, together with partial-specifiedness-aware well-behavedness that supports compositional reasoning and ensures update preservation.
In addition, we demonstrate the utility of the proposed system through examples.

\end{abstract}

\section{Introduction}
\label{sec:introduction}
\label{sec:intro}

Bidirectional transformations provide mechanisms to propagate updates between multiple data representations by translating changes in the transformed result back to the original form. This allows for synchronization across various data structures and has numerous applications, including classical view updating~\cite{BancilhonS81,Hegner90,BohannonPV06,HornPC18}, co-development of printers and parsers~\cite{MaWa13,ZhuK0SH15,ZhuZK0SH16,MatsudaW18haskell,XieSH25},
spreadsheets that support edits to the computed formula results~\cite{CunhaFMPS12,MacedoPSC14,WilliamsG22},
reflexive test case generators that can recover choices from a generated value~\cite{GoldsteinFWP23},
live programming~\cite{ZhangH22,ZhangXGHZH24,MayerKC18},
and model synchronization in model-driven engineering~\cite{Stevens07,YuLHHKM12,XiongLHZTM07}.

Despite their usefulness, developing bidirectional transformations requires effort because programmers must be aware of behavior in both directions. Moreover, such transformations are often expected to satisfy certain properties to be considered ``bidirectional''. Accordingly, many programming languages and frameworks have been proposed to facilitate them~\cite{FosterGMPS05,FGMPS07,BarbosaCFGP10,HofmannPW11,BohannonFPPS08,FosterPP08,HuMT04,MuHT04aplas,MatsudaHNHT07,MatsudaW15,MaWa18esop,MaWa15,MaWa18jfp,HidakaHIKMN10,Voigtlander09bff,VoigtlanderHMW10,VoigtlanderHMW13,KoZH16,HuK16}.
Among these systems, the most prominent is the lens framework~\cite{FosterGMPS05,FGMPS07}.

In the lens framework, a bidirectional transformation---or lens---consists of two functions: $\GET \in  S \to V$ maps a \emph{source} state $s \in S$ to its \emph{view} state $\GET \;s \in V$, and $\PUT \in S \times V \to S$ combines $s$ and an updated view $v'$ to return an updated source, $\PUT  \; (s, v')$.
Since $\GET$ is not always injective, $\PUT$ requires the original source $s$ to reconstruct the information lost during the $\GET$ transformation. 
A lens, a $\GET/\PUT$ pair, is called well-behaved (``bidirectional'') if it satisfies the following two~properties.\footnote{
We use the terminology used in Bancilhon and Spyratos~\cite{BancilhonS81}. Foster \etal ~\cite{FosterGMPS05,FGMPS07} called the acceptability and consistency laws \textsc{GetPut} and \textsc{PutGet}, respectively.
}
\begin{align*}
   \PUT \; (s,v') = s' & \; \Longrightarrow \; \GET \; s' = v' \tag{consistency}\\
   \GET \; s = v & \; \Longrightarrow \; \PUT \; (s, v) = s \tag{acceptability}
\end{align*}
Intuitively, consistency requires that any updates to the view are preserved by $\PUT$-then-$\GET$. 
On the other hand, acceptability ensures that if there is no update to the view, then there is no update to the source. 
The lens framework is outstanding for its \emph{compositional}, \emph{correct-by-construction} approach, allowing programmers to build complex, well-behaved lenses by composing simpler ones and language designers to extend the system without breaking the guarantee. 

Since their first appearance, lenses have been extended in various ways, but several research challenges remain. One key problem is the handling of multiple views. When multiple views share the same source, and a user updates one of them, we expect the other views to be adjusted accordingly while preserving the original change in the view that was directly modified.
However, allowing this behavior breaks traditional well-behavedness properties, particularly \emph{consistency}. 
Specifically, the view state $v'$ that a user provides may differ from the final state $\GET \; (\PUT \; (s,v'))$ after synchronization, 
because an update (involved in $v'$) to one view will be propagated to the other view via the shared source. 
More generally, when multiple views are updated simultaneously, 
these updates are merged and propagated through the source, 
meaning no single updated view is guaranteed to remain unchanged.
The technical challenges are: 
\begin{fullversionblock}
\begin{itemize}
\item characterizations of update preservation (for \emph{consistency}) and of no update (for \emph{acceptability}), and 
\item an appropriate \emph{compositional} notion of well-behavedness based on the characterizations.
\end{itemize} 
\end{fullversionblock}

To the best of our knowledge, existing approaches either rely on global (non-compositional) properties or adopt weaker properties without update preservation. For example, Hu~\etal~\cite{HuMT04} and Mu~\etal~\cite{MuHT04aplas} propose the duplication lens to support multiple views with the desired behavior, but they abandon the preservation of updates (consistency). As a remedy, Hidaka~\etal~\cite{HidakaHIKMN10} propose a weaker consistency property called \textsc{WPutGet}, where the modified and final views can differ, but both must result in the same update to the original source; this neither preserves the very update the user provided, nor is it compositional.
As far as we are aware, using existing generalizations/extensions to lenses does not resolve this issue.
Many of them~(\eg, \cite{DiskinXC11,AhmanU17,HofmannPW11,HofmannPW12,Abou-SalehCGMS16}) require the updated view to be identical to the view of the updated source, which is too strong to accommodate the above behavior of multiple views. 
Some categorical generalizations~\cite{Boisseau20,riley2018categoriesoptics} assume a form of the \textsc{PutPut} law~\cite{FGMPS07}, which intuitively states that $\PUT$ must preserve update composition. 
However, in the literature of view updating, this law is considered too strong to permit practical transformations~(see, \eg, \cite{FGMPS07,Keller87}).
To specify whether an update is preserved in the presence of multiple views sharing a source, we need a clear notion
to represent the intentions underlying the user's updates. 
Consider the simplest situation where a source is simply copied into two views. For example, if the source state is $5$, the corresponding view is $(5,5)$. 
Suppose that this view is changed to $(7, 5)$. 
If the user's intention is just to change the first view from $5$ to $7$, without caring about the second, we can safely propagate the update to obtain the updated source $7$ whose view is $(7, 7)$, where the user's intention is preserved. 
In contrast, if the user's intention is to change the pair of views to $(7, 5)$, there is no way to propagate the change to the source while preserving the user's intention. 
However, if we represent the updated state as $(7, 5)$, the intention is ambiguous. 
In general, we can have simultaneous updates on multiple views with independent intentions. 
In such cases, we try to merge these intentions into one.

This observation naturally suggests representing these update intentions by using partial orders, where comparing two update intentions $u_1 \le u_2$ means that $u_1$ is subsumed by, or preserved in $u_2$.
For example, in the previous example, we can define $\GO \le n$ for any $n$, where $\GO$ represents an unspecified number, and $(7, \GO) \le (7, 7)$ indicates that the intention of updating only the first copy to $7$ is preserved in the view $(7,7)$.
Additionally, partial orders provide a natural notion of merging two update intentions: the least upper bound or the join ($\vee$). The join operator is partial in general, but this partiality is rather desirable to model conflicting updates: recall that, if we intend to update one copy of a source to $7$ and another to $5$, no source update can reflect these conflicting intentions. 

A similar idea appears in conflict-free replicated datatypes (CRDTs)~\cite{CRDT}, which are used to synchronize distributed replicas that are updated in an independent and concurrent manner without coordination. 
In state-based CRDTs, replica states form a join-semilattice, where the $(\vee)$ operator is used to merge updates. As long as updates make states greater than or equal to the original states, the commutative, associative, and idempotent properties of the join guarantee that every replica reaches a consistent state without rollbacks. 
Sending out whole states appears impractical, but it is known that it suffices to send {delta states} $u$, where an update is described as $s \vee u$~\cite{delta-CRDT}.
Although requiring join-semilattices and monotonic updates is too strong for our purposes in general, this connection suggests two advantages of handling partially ordered domains. 
First, our bidirectional transformations can now handle CRDTs to enlarge potential applications. 
Second, we can borrow some ideas from concrete CRDTs to design update intentions for concrete bidirectional transformation examples. 

Equipped with partial orders to represent update intentions, a remaining challenge is to establish an appropriate notion of ordering-aware well-behavedness that is suitable for compositional reasoning. %
The well-behavedness should be independent of concrete representations of \pstates, to support both simple \pstates that just allow complete unspecifiedness and \pstates tailored to concrete datatypes like CRDTs.

In this paper, we propose a novel bidirectional transformation framework, which we call {\our} lenses. Its key feature lies in lenses that can operate on partially-ordered update intentions, which we call \pstates. 
Leveraging the partial ordering, our framework provides compositional well-behavedness that guarantees preservation of updates across multiple views sharing a source, independently of concrete representations of \pstates.
Specifically, we make the following contributions.

\begin{itemize}
 \item We illustrate our key idea with a concrete scenario (\cref{sec:scenarios}). After motivating the source-sharing problem with it, we observe  a limitation of classical lenses: the lack of a domain capable of expressing the user's intent, namely, the intent of what an update should finally result in. We then demonstrate how having \pstates addresses this issue.
 \item We formalize the {\our{} lens} framework, which incorporates compositional well-behavedness concerning partial specifiedness (\cref{sec:absence-aware-lenses}).
 This is the first lens framework that achieves all of: (1) view-to-view update propagation in multiple views via a shared source, (2) guaranteed preservation of user's updates, (3) compositional reasoning, and (4) interoperability with the classical lens framework~\cite{FGMPS07,FosterGMPS05}. 
 \item We demonstrate the utility of the proposed framework by revisiting the scenario and informal solution from \cref{sec:scenarios}, showing how our  framework can model the scenario through the composition of concrete lenses  (\cref{sec:revisiting}).
 \item We give a recipe to encode updates into \pstates, demonstrating that {\our} lenses are hybrids of state-based systems and operation-based systems~\cite{HofmannPW12,DiskinXC11,AhmanU17} (\cref{sec:ours-as-operation-based-system}).
\end{itemize}
Finally, we discuss related work (\cref{sec:related-work}) and conclude the paper (\cref{sec:conclusion}).
{We include some proofs in the appendix (\cref{sec:proofs}).}{}
We also provide mechanized proofs in Agda\footnote{\url{https://github.com/kztk-m/ps-lenses-agda}} and a prototype implementation in Haskell.\footnote{\url{https://github.com/kztk-m/ps-lenses-hs}}

\section{Motivating Scenario: Shared-Source Problem}
\label{sec:scenarios}

In this section, we will provide an overview of our approach using a motivating example. 

\newcommand{\Changed}[1]{\textbf{#1}}

\newcommand{\TaskA}{{Buy milk}\xspace}
\newcommand{\TaskB}{{Walk dog}\xspace}
\newcommand{\TaskC}{{Jog}\xspace}
\newcommand{\TaskD}{{Buy egg}\xspace}
\newcommand{\TaskE}{{Stretch}\xspace}

\newcommand{\OSourceVar}{s_\mathrm{tl}}
\newcommand{\OViewOGVar}{v_\mathrm{og}}
\newcommand{\OViewDTVar}{v_\mathrm{dt}}

\newcommand{\OrigSource}{%
    \{%
        \bbt \str{Brown} :\\%
        \phantom{\texttt{X}}\{ \bbt  \str{001} : \{ \bbt \str{Done} :\texttt{false},\\ \str{Name} : \TaskA \},\ee  \\%
                     \str{002} : \{ \bbt \str{Done} :\texttt{true},\\ \str{Name} : \TaskB \}\}, \ee %
               \ee \\%
            \str{Smith} :\\%
            \phantom{\texttt{X}}\{ \str{001} : \{ \bbt \str{Done} : \texttt{false}, \\ \str{Name} : \TaskC \}\}\}\ee%
        \ee%
}%
\newcommand{\BrownView}{%
\{ \bbt  \str{001} : \{ \bbt \str{Done} :\texttt{false},\\ \str{Name} : \TaskA \},\ee  \\%
         \str{002} : \{ \bbt \str{Done} :\texttt{true},\\ \str{Name} : \TaskB \}\} \ee %
\ee%
}%
\newcommand{\SmithView}{%
\{ \str{001} : \{ \bbt \str{Done} : \texttt{false}, \\ \str{Name} : \TaskC \}\ee%
}

\newcommand{\SmithViewU}{%
\{ \str{001} : \{ \bbt \str{Done} : \texttt{true}, \\ \str{Name} : \TaskC \}\ee%
}
\newcommand{\BrownViewU}{%
\{ \bbt  \str{001} : \{ \bbt \str{Done} :\texttt{true},\\ \str{Name} : \TaskA \},\ee  \\%
         \str{002} : \{ \bbt \str{Done} :\texttt{true},\\ \str{Name} : \TaskB \}\} \ee %
\ee%
}%

\newcommand{\BrownSourceU}{%
    \{%
        \bbt \str{Brown} :\\%
        \phantom{\texttt{X}}\{ \bbt  \str{001} : \{ \bbt \str{Done} :\texttt{true},\\ \str{Name} : \TaskA \},\ee  \\%
                     \str{002} : \{ \bbt \str{Done} :\texttt{true},\\ \str{Name} : \TaskB \}\}, \ee %
               \ee \\%
            \str{Smith} :\\%
            \phantom{\texttt{X}}\{ \str{001} : \{ \bbt \str{Done} : \texttt{false}, \\ \str{Name} : \TaskC \}\}\}\ee%
        \ee%
}%
\newcommand{\SmithSourceU}{%
    \{%
        \bbt \str{Brown} :\\%
        \phantom{\texttt{X}}  \{ \bbt  \str{001} : \{ \bbt \str{Done} :\texttt{false},\\ \str{Name} : \TaskA \},\ee  \\%
                     \str{002} : \{ \bbt \str{Done} :\texttt{true},\\ \str{Name} : \TaskB \}\}, \ee %
               \ee \\%
               \phantom{\texttt{X}}\str{Smith} :\\%
            \{ \str{001} : \{ \bbt \str{Done} : \texttt{true}, \\ \str{Name} : \TaskC \}\}\}\ee%
        \ee%
}%
\newcommand{\SourceU}{%
    \{%
        \bbt \str{Brown} :\\%
            \phantom{\texttt{X}}\{ \bbt  \str{001} : \{ \bbt \str{Done} :\texttt{true},\\ \str{Name} : \TaskA \},\ee  \\%
                     \str{002} : \{ \bbt \str{Done} :\texttt{true},\\ \str{Name} : \TaskB \}\}, \ee %
               \ee \\%
            \str{Smith} :\\%
            \phantom{\texttt{X}}\{ \str{001} : \{ \bbt \str{Done} : \texttt{true}, \\ \str{Name} : \TaskC \}\}\}\ee%
        \ee%
}%

\newcommand{\BrownSourceUSpec}{%
 \bbt
    {\oplus}\{%
        \bbt \str{Brown} : \\%
            \phantom{\texttt{X}}\{ \bbt  \str{001} : \{ \bbt \str{Done} :\texttt{true},\\ \str{Name} : \TaskA \},\ee  \\%
                     \str{002} : \{ \bbt \str{Done} :\texttt{true},\\ \str{Name} : \TaskB \}\}\} \ee %
               \ee
        \ee%
    \\/{\ominus}\emptyset%
 \ee
}%
\newcommand{\SmithSourceUSpec}{%
 \bbt
    {\oplus}\{%
        \bbt \str{Smith} :\\%
         \phantom{\texttt{X}}\{ \str{001} : \{ \bbt \str{Done} : \texttt{true}, \\ \str{Name} : \TaskC \}\}\}/{\ominus}\emptyset\ee%
        \ee%
 \ee
}%
\newcommand{\SourceUSpec}{%
    {\oplus}\{%
        \bbt \str{Brown} :\\%
        \phantom{\texttt{X}}\{ \bbt  \str{001} : \{ \bbt \str{Done} :\texttt{true},\\ \str{Name} : \TaskA \},\ee  \\%
                     \str{002} : \{ \bbt \str{Done} :\texttt{true},\\ \str{Name} : \TaskB \}\}, \ee %
               \ee \\%
            \str{Smith} :\\%
            \phantom{\texttt{X}}\{ \str{001} : \{ \bbt \str{Done} : \texttt{true}, \\ \str{Name} : \TaskC \}\}\}/{\ominus}\emptyset\ee%
        \ee%
}%

\subsection{A Motivating Scenario}
\label{sec:motivating-example}

\tikzset{tbox/.style={shape=rectangle, inner sep=2pt, fill={rgb,255:red,245; green, 245; blue,247}}}

\newcommand{\TODAY}{Apr 1}
\newcommand{\NEXTDAY}{Apr 2}

\newcommand{\filterOnGoing}{\ensuremath{\var{filter}_{\mathrm{\neg{}Done}}}\xspace}
\newcommand{\filterToday}{\ensuremath{\var{filter}_{\mathrm{\TODAY}}}\xspace}

\newcommand{\OrigTable}{
\begin{tabular}{@{}l|lll@{}}
ID & Done & Name & Due \\\hline
\texttt{001} & \FF & \TaskA & \NEXTDAY\\
\texttt{002} & \TT & \TaskB & \TODAY\\
\texttt{003} & \FF & \TaskC & \TODAY  
\end{tabular}}

\newcommand{\OrigFilteredOGTable}{
\begin{tabular}{@{}l|lll@{}}
ID & Done & Name & Due \\\hline
\texttt{001} & \FF & \TaskA & \NEXTDAY\\
\texttt{003} & \FF & \TaskC & \TODAY
\end{tabular}}

\newcommand{\OrigFilteredDTTable}{
\begin{tabular}{@{}l|lll@{}}
ID & Done & Name & Due \\\hline
\texttt{002} & \TT & \TaskB & \TODAY\\
\texttt{003} & \FF & \TaskC & \TODAY  
\end{tabular}}

\begin{figure}[t]
\centering
\let\varsize\footnotesize%
\let\tbsize\tinier%
\let\funsize\tinier%
\begin{tikzpicture}[node distance=1.3cm]
\node[tbox] (s) {\tbsize\OrigTable};
\node[tbox] (s1) [right=0.8cm of s.north east, yshift= 1mm] {\tbsize\OrigTable}; 
\node[tbox] (s2) [right=0.8cm of s.south east, yshift=-1mm] {\tbsize\OrigTable}; 
\node[tbox] (v1) [right=of s1] {\tbsize\OrigFilteredOGTable};
\node[tbox] (v2) [right=of s2] {\tbsize\OrigFilteredDTTable};
\draw[-{Stealth}, rounded corners, line width=1pt] let \p1 = ( $(s.east)!0.5!(s1.west)$ ) in
      (s.east) -- (s.east -| \p1) -- (\p1 |- s1.west) -- (s1.west);
\draw[-{Stealth}, rounded corners, line width=1pt] let \p1 = ( $(s.east)!0.5!(s2.west)$ ) in
      (s.east) -- (s.east -| \p1) -- (\p1 |- s2.west) -- (s2.west);
\path let \p1 = ( $(s.east)!0.5!(s1.west)$ ) in
      (\p1 |- s1.west) node[above] {\funsize$\DUP$} ;
\draw[-{Stealth}, line width=1pt] (s1) -- node[above] {\funsize$\filterOnGoing$} (v1) ;       
\draw[-{Stealth}, line width=1pt] (s2) -- node[above] {\funsize$\filterToday$} (v2) ;       
\end{tikzpicture}
\caption{Forward (\GET) behavior of composition of lenses $\DUP$ and $\filterOnGoing$/$\filterToday$}
\label{fig:running-example-get}
\end{figure}

Consider managing tasks (to-dos) using two views: one (\VName{ongoing}) for all ongoing tasks and the other (\VName{today}) for all tasks due today.
We assume that these tasks may be updated individually, maybe because the views correspond to different panes in the same application or because these tasks are shared with others. This example models a situation where bidirectional transformations are used to extract (and transform) data for application components to manage. Such an application is a common introductory example  %
in Web application programming.\footnote{See, for example, \url{https://todomvc.com/}, though with different views.} We could go further to extract smaller pieces of data for smaller components (such as a task name for a text box), which would be preferable, but for simplicity we here consider these two, coarser-grained views. 

\ifFullVersion
\else 
\begin{wrapfigure}[10]{r}{5.1cm}%
\ifdim\pagetotal=0pt \else\vspace{-\intextsep}\fi%
\bgroup%
\ifHighlightVersionDiff\color{diffcolor}\fi %
\centering
\(
\begin{aligned}    
\OSourceVar &= \left(
\text{\myfootnotesize\OrigTable}
\right)
\\
\OViewOGVar &= \left(
\text{\myfootnotesize\OrigFilteredOGTable}\right) 
\\
\OViewDTVar &= \left(\text{\myfootnotesize\OrigFilteredDTTable}\right)
\end{aligned}
\)
\egroup
\end{wrapfigure}
\fi

For example, for the set $\OSourceVar$ of whole tasks {below}, 
\begin{fullversionblock}
\[
\OSourceVar = \left(
\text{\myfootnotesize\begin{tabular}{l|lll}
ID & Done & Name & Due \\\hline
\texttt{001} & \FF & \TaskA & \NEXTDAY \\
\texttt{002} & \TT  & \TaskB & \TODAY \\
\texttt{003} & \FF & \TaskC & \TODAY 
\end{tabular}}
\right)
\]    
\end{fullversionblock}
its corresponding views are $\OViewOGVar$ and $\OViewDTVar$ {(assuming today is \TODAY):
\[
\OViewOGVar = \left(
\text{\myfootnotesize\begin{tabular}{l|lll}
ID & Done & Name & Due \\\hline
\texttt{001} & \FF & \TaskA & \NEXTDAY \\
\texttt{003} & \FF & \TaskC & \TODAY 
\end{tabular}
}\right) 
\qquad 
\OViewDTVar = \left(
\text{\myfootnotesize\begin{tabular}{l|lll}
ID & Done & Name & Due \\\hline
\texttt{002} & \TT  & \TaskB & \TODAY \\
\texttt{003} & \FF  & \TaskC & \TODAY 
\end{tabular}}
\right)
\]
}{}
(We use table-like notations solely for readability; we do not make any assumptions on their actual representations at this point.)

Then, we consider connecting the source with the two views by bidirectional transformations. 
Suppose that the connections from the whole source to these views are given by $\filterOnGoing$ and $\filterToday$. 
Instead of considering them individually, we can bundle them by using the duplication lens~\cite{MuHT04aplas,HuMT04}.
The entire transformation is hence given as a composition of $\DUP$ and $\filterOnGoing$/$\filterToday$, whose forward behavior is depicted in \cref{fig:running-example-get}. 
In the forward direction, $\DUP$ simply copies the source, so that lenses to be composed can work on their copies. In the backward direction, $\DUP$ merges updates on the copies---we leave the concrete definition for now, as how to provide a suitable one for compositional reasoning is a subject of discussion in this section. Theoretically, $\DUP$ enables us to focus on a single lens instead of considering interactions among lenses. Practically, thanks to the internalization, we can introduce sharing dynamically in the transformation, allowing us to compose lenses involving sharing. %
A $\DUP$-like operation is particularly useful in giving a compositional semantics to (non-linear) bidirectional programming languages~\cite{MaWa18esop,HidakaHIKMN10}.

\subsection{Issues}

\ifFullVersion
\else 
\begin{wrapfigure}[4]{r}{5.2cm}
\ifdim\pagetotal=0pt \else\vspace{-1\intextsep}\fi%
\centering\(\begin{aligned}
\OViewOGVar' &= \left(\text{\myfootnotesize
\begin{tabular}{l|lll}
ID & Done & Name & Due \\\hline
\texttt{001} & \FF & \TaskA & \NEXTDAY \\
\texttt{003} & \FF & \TaskC & \TODAY \\
\texttt{004} & \FF & \TaskD & \TODAY 
\end{tabular}
}\right)
\end{aligned}
\)
\end{wrapfigure}
\fi 

Consider adding a new task ``\TaskD'' to the \VName{ongoing} view, {as below. 
\[
\OViewOGVar' = \left(\text{\myfootnotesize
\begin{tabular}{l|lll}
ID & Done & Name & Due \\\hline
\texttt{001} & \FF & \TaskA & \NEXTDAY \\
\texttt{003} & \FF & \TaskC & \TODAY \\
\texttt{004} & \FF & \TaskD & \TODAY 
\end{tabular}
}\right)
\]}{}
After propagating this update by the backward transformation ($\PUT$) and by the subsequent forward transformation ($\GET$), the added tuple is expected to appear in the original source and the \VName{today} view as follows:
\[
\OSourceVar' = \left(
\text{\myfootnotesize\begin{tabular}{l|lll}
ID & Done & Name & Due \\\hline
\texttt{001} & \FF & \TaskA & \NEXTDAY \\
\texttt{002} & \TT  & \TaskB & \TODAY \\
\texttt{003} & \FF & \TaskC & \TODAY \\
\texttt{004} & \FF & \TaskD & \TODAY 
\end{tabular}}
\right)
\qquad 
\OViewDTVar' = \left(
\text{\myfootnotesize\begin{tabular}{l|lll}
ID & Done & Name & Due \\\hline
\texttt{002} & \TT  & \TaskB & \TODAY \\
\texttt{003} & \FF  & \TaskC & \TODAY\\ 
\texttt{004} & \FF & \TaskD & \TODAY 
\end{tabular}}
\right)\text{.}
\]

To achieve this behavior, the backward behavior of $\DUP$ must take $(\OSourceVar', \OSourceVar)$ in addition to the original source $\OSourceVar$ to return $\OSourceVar'$. 
Existing $\DUP$ lenses achieve this by detecting which copies are updated by comparing them with the original source~\cite{HuMT04} or by marking explicitly which view is updated~\cite{MuHT04aplas}. However, this approach has two issues. First, this particular behavior violates the consistency property~\cite{FGMPS07,BancilhonS81} presented in \cref{sec:intro}, which states \emph{preservation of updates}; we updated the views to $(\OViewOGVar', \OViewDTVar)$ but finally obtained $(\OViewOGVar', \OViewDTVar')$ after $\PUT$-then-$\GET$.
To the best of our knowledge, no notion of update preservation exists that is compositional and allows such view-to-view update propagation via a shared source. 
Second, the approach does not scale to cases where two views are updated simultaneously. 
Simultaneous updating introduces a further difficulty: even when focusing on a single view, the consistency property no longer holds in general. 
Even without simultaneous updating, stating the consistency by focusing on a single view is unsatisfactory because it complicates the reasoning: we need to perform aliasing analysis to prevent copies from being mixed up in transformations, as otherwise we cannot project ``a view'' from the final result.

\subsection{Approach: \OUR Lenses}
\label{sec:approach-overview}

The above issues stem from the fact that mere states are not expressive enough to represent the user's intentions. For example, if we can state that what is changed in the views $(\OViewOGVar', \OViewDTVar)$ is the insertion of the task ID \texttt{004} to the first view, we can state the update preservation as the inserted task is present after the synchronized state $(\OViewOGVar', \OViewDTVar')$. 
To state update preservation, such intentions must be compared, and to support simultaneous updating, they must come with a merge operation. This suggests that such intentions must be partially-ordered. 

We call these intentions \emph{\pstates} and allow our lenses to handle them in addition to ordinary \emph{\fstateHead} states. One may think that \pstates are hybrids of states and updates (also called deltas~\cite{DiskinXC11}, edits~\cite{HofmannPW12} or operations~\cite{Meertens98,CRDT}). 
Like updates, they can express finer intentions---for example, distinguishing an insertion of a single task from the entire state change---while, as states, they allow update preservation to be stated simply by comparing them using the partial order. 
A similar idea has been adopted in delta-state CRDTs~\cite{delta-CRDT}, which are also partially-ordered and mix advantages of deltas and states, where the merge operation is guaranteed to be total---this is not the case for us as it is too strong for our purposes. 
Unlike CRDTs, where distributed replicas are updated asynchronously without coordination, the execution of $\PUT$ is often coordinated and synchronous, which makes failing a feasible option, compared with silent resolution of competing updates. 

\newcommand{\DStateOGVar}{w_\mathrm{og}}
\newcommand{\DStateDTVar}{w_\mathrm{dt}}

\newcommand{\DTableOG}{\begin{tabular}{@{}l|llll@{}}
    ID & Done & Name & Due & \(\delta\) \\\hline 
    \texttt{004} & \FF & \TaskD & \TODAY & {$+$} 
  \end{tabular}}

\newcommand{\DTableDT}{\begin{tabular}{@{}l|llll@{}}
    ID & Done & Name & Due & \(\delta\) \\\hline 
    \texttt{002} &      &     &  & {$-$} \\
    \texttt{003} & \FF & \TaskE & \TODAY & {$+$} 
  \end{tabular}}

\newcommand{\DTableM}{\begin{tabular}{@{}l|llll@{}}
    ID & Done & Name & Due & \(\delta\) \\\hline 
    \texttt{002} &     &        &       & {$-$} \\
    \texttt{003} & \FF & \TaskE & \TODAY & {$+$} \\
    \texttt{004} & \FF & \TaskD & \TODAY & {$+$} 
  \end{tabular}}

Let us go back to the motivating example in \cref{sec:motivating-example}. Leaving the formal definition of \pstates used for this example to \cref{sec:revisiting}, here we will illustrate its intuitive behavior for \pstates. When a user adds a task with ID \texttt{004} ``\TaskD'' due {\TODAY} to the \VName{ongoing} view and leaves the \VName{today} view, if the user's intention underlying the update is to keep the added task in the \VName{ongoing} view, the intention is expressed as the pair $(\DStateOGVar, \Omega)$ with 
\[
 \DStateOGVar = \left(\text{\myfootnotesize\DTableOG}\right)\text{.}
\]
Here, for visual simplicity, we represent (strictly) \pstates as tables with an extra field $\delta$ for update marks. %
The $(+)$ mark indicates the intention that the corresponding tuple must be present; \ie, it requests upsertion of the marked tuples. We also have the $(-)$ mark that indicates that the corresponding tuple must not be present; 
$(-)$-marked tuples only hold IDs because the information is sufficient for deletion. 
Formally, these intentions are modeled by ordering between a \pstate and a \fstate.
\Spstates are ordered by the subset inclusion, and 
the least element is denoted by $\Omega$,%
\footnote{To avoid confusion with $\bot$ often used to denote undefinedness of a partial function, we use $\Omega$ to denote the least element (if exists) in \pstates.}  
which corresponds to the empty table and intuitively means ``anything''. 

The backward behaviors of $\filterOnGoing$ and $\filterToday$ simply pass $\DStateOGVar$ and $\Omega$ through. They are then merged by $\DUP$ using $\vee$ as $\DStateOGVar \vee \Omega = \DStateOGVar$; recall that $\Omega$ is the least element, which is the unit of $\vee$. This result indicates that the $(+)$-marked task in $\DStateOGVar$ should be inserted into $\OSourceVar$ (which results in $\OSourceVar'$) to accommodate the view update $(\DStateOGVar, \Omega)$.

We may end the backward transformation here, assuming some external process to reflect $\DStateOGVar$ into $\OSourceVar$, by leveraging the update aspect of a \pstate $\DStateOGVar$.
Instead, similarly to $\DUP$ that internalizes the source sharing, we also internalize this updating process as a lens in our framework. This not only simplifies our formalization but also enables a compositional design of such update reflection. 
Specifically, we use a lens called a \emph{\initiator} whose $\GET$ embeds \fstates into \pstates and whose $\PUT$ (tries to) reflect a \pstate into the original \fstate, implementing the semantics of a \pstate as an update. 
Let us write $\TaskInit$ for a particular \initiator for this example, whose $\PUT$ is designed to take, for example, $(\OSourceVar, \DStateOGVar)$ to return $\OSourceVar'$. 
In general, its $\PUT$ for $(s, w)$ updates or inserts tasks marked $(+)$ in $w$ into $s$, depending on whether a task with the same ID already exists in $s$ (\ie, its $\PUT$ upserts the tasks marked $(+)$ in $w$ into $s$), and then deletes tasks marked by $(-)$. 
This behavior of the $\PUT$ of $\TaskInit$ respects the above-mentioned meaning of \pstates; formally, its $\PUT$ is defined so that $w \le s'$ holds for $s'$ with $\PUT \; (s, w) = s'$---in particular, $\DStateOGVar \le \OSourceVar'$.
Now the source is updated to $\OSourceVar'$. 
The views of $\OSourceVar'$ are $(\OViewOGVar', \OViewDTVar')$. 
We have $\DStateOGVar \le \OViewOGVar'$ for the reason we have explained earlier, and $\Omega \le \OViewDTVar'$ as $\Omega$ is the least element. 
Then, we have $(\DStateOGVar, \Omega) \le (\OViewOGVar', \OViewDTVar')$ by point-wise ordering, stating that the update intention $(\DStateOGVar, \Omega)$ is in fact preserved. 

\newcommand{\SourceTableUU}{\begin{tabular}{@{}l|lll@{}}
ID & Done & Name & Due \\\hline
\texttt{001} & \FF & \TaskA & \NEXTDAY \\
\texttt{003} & \FF & \TaskE & \TODAY \\
\texttt{004} & \FF & \TaskD & \TODAY 
\end{tabular}}

\newcommand{\ViewOGTableUU}{\begin{tabular}{@{}l|lll@{}}
ID & Done & Name & Due \\\hline
\texttt{001} & \FF & \TaskA & \NEXTDAY \\
\texttt{003} & \FF & \TaskE & \TODAY \\
\texttt{004} & \FF & \TaskD & \TODAY 
\end{tabular}}

\newcommand{\ViewDTTableUU}{\begin{tabular}{@{}l|lll@{}}
ID & Done & Name & Due \\\hline
\texttt{003} & \FF & \TaskE & \TODAY \\
\texttt{004} & \FF & \TaskD & \TODAY 
\end{tabular}}

\begin{figure}[t]
\let\tbsize\tiny
\begin{subfigure}[t]{1\textwidth}\centering
    \scalebox{0.85}{\begin{tikzpicture}
    \node[tbox] (s0) [left=0.8cm of s.west] {\tbsize\SourceTableUU};
    \node[tbox] (s) {\tbsize\DTableM};
    \node[tbox] (s1) [right=0.6cm of s.north east, yshift=-0.8mm] {\tbsize\DTableOG}; 
    \node[tbox] (s2) [right=0.6cm of s.south east, yshift=+0.8mm] {\tbsize\DTableDT}; 
    \node[tbox] (v1) [on grid, right=4.2cm of s1] {\tbsize\DTableOG};
    \node[tbox] (v2) [on grid, right=4.2cm of s2] {\tbsize\DTableDT};
    \draw[{Stealth}-, line width=1pt] (s0) -- node[above] {\tiny$\TaskInit$} (s);
    \draw[{Stealth}-, rounded corners, line width=1pt] let \p1 = ( $(s.east)!0.7!(s1.west)$ ) in
      (s.east) -- (s.east -| \p1) -- (\p1 |- s1.west) -- (s1.west);
\draw[{Stealth}-, rounded corners, line width=1pt] let \p1 = ( $(s.east)!0.7!(s2.west)$ ) in
      (s.east) -- (s.east -| \p1) -- (\p1 |- s2.west) -- (s2.west);
\path let \p1 = ( $(s.east)!0.5!(s1.west)$ ) in
      (\p1 |- s1.west) node[above] {\tiny$\DUP$} ;
\draw[{Stealth}-, line width=1pt] (s1) -- node[above] {\tiny$\filterOnGoing$} (v1) ;       
\draw[{Stealth}-, line width=1pt] (s2) -- node[above] {\tiny$\filterToday$} (v2) ;       
    \end{tikzpicture}}\vspace{-2pt}
    \caption{Backward ($\PUT$), reusing original sources from \cref{fig:running-example-get}; all involved lenses are lawful.}
     \label{fig:running-example-our-put}
     \medskip
\end{subfigure}
\begin{subfigure}[t]{1\textwidth}\centering
    \scalebox{0.85}{\begin{tikzpicture}
    \node[tbox] (s0) [left=0.8cm of s.west] {\tbsize\SourceTableUU};
    \node[tbox] (s) {\tbsize\SourceTableUU};
    \node[tbox] (s1) [right=0.7cm of s.north east, yshift= 1.0mm] {\tbsize\SourceTableUU}; 
    \node[tbox] (s2) [right=0.7cm of s.south east, yshift=-1.0mm] {\tbsize\SourceTableUU}; 
    \node[tbox] (v1) [on grid, right=4.2cm of s1] {\tbsize\ViewOGTableUU};
    \node[tbox] (v2) [on grid, right=4.2cm of s2] {\tbsize\ViewDTTableUU};
    \draw[-{Stealth}, line width=1pt] (s0) -- node[above] {\tiny$\TaskInit$} (s);
    \draw[-{Stealth}, rounded corners, line width=1pt] let \p1 = ( $(s.east)!0.4!(s1.west)$ ) in
      (s.east) -- (s.east -| \p1) -- (\p1 |- s1.west) -- (s1.west);
\draw[-{Stealth}, rounded corners, line width=1pt] let \p1 = ( $(s.east)!0.4!(s2.west)$ ) in
      (s.east) -- (s.east -| \p1) -- (\p1 |- s2.west) -- (s2.west);
\path let \p1 = ( $(s.east)!0.5!(s1.west)$ ) in
      (\p1 |- s1.west) node[above] {\tiny$\DUP$} ;
\draw[-{Stealth}, line width=1pt] (s1) -- node[above] {\tiny$\filterOnGoing$} (v1) ;       
\draw[-{Stealth}, line width=1pt] (s2) -- node[above] {\tiny$\filterToday$} (v2) ;       
    \end{tikzpicture}}\vspace{-2pt}
            \caption{Forward ($\GET$) after the Backward Execution ($\PUT$) in \cref{fig:running-example-our-put}}
        \label{fig:running-example-our-get}
\end{subfigure}
    \caption{Composition of \our lenses $\DUP$ and $\filterOnGoing$/$\filterToday$ for simultaneous updates}
    \label{fig:running-example-our}
\end{figure}

This approach extends easily to simultaneous updates. 
Suppose that the user also deletes the task with ID \texttt{002} and changes the name of the task with ID \texttt{003} to ``\TaskE'' on the \VName{today} view, simultaneously with the update $\DStateOGVar$ on the \VName{ongoing} view. Suppose also that the underlying intention of the update is exactly the deletion and the task name change. Then, we can represent the user's intention as $(\DStateOGVar, \DStateDTVar)$ with
\[
\DStateDTVar = \left(\text{\myfootnotesize\DTableDT}\right)\text{.}
\]
Again, the backward behaviors of $\filterOnGoing$ and $\filterToday$ do nothing interesting and return $(\DStateOGVar, \DStateDTVar)$ intact. Then, the $\PUT$ of $\DUP$ merges the two updates as:
\[
w_\mathrm{merged} = \DStateOGVar \vee \DStateDTVar = 
\left(\text{\myfootnotesize\DTableM}\right)
\]
After that, the $\PUT$ of $\TaskInit$ reflects $w_\mathrm{merged}$ into $\OSourceVar$, which results in: 
\[
 \OSourceVar'' = \left(
\text{\myfootnotesize\SourceTableUU}
\right)
\]
Accordingly, by the forward transformation, the views are changed to:
\[
\OViewOGVar'' = \left(\text{\myfootnotesize\ViewOGTableUU}\right)
\quad 
\OViewDTVar'' = \left(
\text{\myfootnotesize\ViewDTTableUU}
\right)
\]
This backward and forward behavior is depicted in \cref{fig:running-example-our}.
Observe again that $\DStateOGVar$ and $\DStateDTVar$ are preserved in $\OViewOGVar''$ and $\OViewDTVar''$, respectively. 

\subsection{Additional Advantage: Fine-Grained Descriptions of Updates}
\label{sec:finer-grained-descriptions-of-updates}

As we mentioned earlier, a \pstate lies in between a (\fstateHead) state and an update. This dual aspect allows our state-based framework to enjoy some advantages of operation-based systems~\cite{DiskinXC11,HofmannPW12,AhmanU17,Meertens98}. Our $\PUT$ can distinguish updates that have the same net effect on states, because \pstates can convey information that is not visible from \fstates. 

Recall that the motivating example models the extraction of data managed in a to-do management application. Thus, it is desirable if the views support operations such as task insertion, task renaming, task completion and task postponing. However, this is known to be difficult in purely state-based systems. For example, in a state-based system like classical lenses~\cite{FGMPS07}, we cannot distinguish task completions from task deletions for the \VName{ongoing} view as their net effect on states is the same; both completion and deletion will remove the task from the ongoing view. Similarly, for the \VName{today} view, we cannot distinguish task postponing from task deletions for the same reason. 
The need to distinguish such updates is a key motivation for operation-based systems~\cite{DiskinXC11,HofmannPW12,Meertens98}, where $\PUT$ transforms updates (or operations) rather than states. 

\newcommand{\Postpone}{\mathrm{P}}

\newcommand{\DTableCompleteOG}{
\begin{tabular}{@{}l|llll@{}}
ID & Done & Name & Due & $\delta$\\\hline 
\texttt{001} &     &        &          & {$-$} \\
\texttt{003} & \TT & \TaskC & \TODAY   & {$\checkmark$}
\end{tabular}%
}

\newcommand{\OrigTableAfterCompleteOG}{%
\begin{tabular}{@{}l|lll@{}}
ID & Done & Name & Due \\\hline
\texttt{002} & \TT  & \TaskB & \TODAY \\
\texttt{003} & \TT & \TaskC & \TODAY 
\end{tabular}%
}

\ifFullVersion 
\begin{wrapfigure}[7]{R}{5cm}
\ifdim\pagetotal=0pt \else\vspace{-0.8\intextsep}\fi
\centering{\myfootnotesize\DTableCompleteOG}
\caption{A \pstate involving completion and deletion}
\label{fig:a-table-involving-completion}
\end{wrapfigure}
\else 
\begin{wrapfigure}[13]{R}{5cm}
\ifdim\pagetotal=0pt \else\vspace{-0.8\intextsep}\fi
\centering{\myfootnotesize\DTableCompleteOG}
\caption{A \pstate involving completion and deletion}
\label{fig:a-table-involving-completion}
\vskip\baselineskip
\centering{\myfootnotesize\OrigTableAfterCompleteOG}
\caption{The update source corresponding to the update in \cref{fig:a-table-involving-completion}}
\label{fig:resulting-table-involving-completion}
\end{wrapfigure}
\fi

\Pstates can also tackle this particular challenge by extending \pstates. %
For the \VName{ongoing} view, or the view of \filterOnGoing, we consider the mark $(\checkmark)$ in addition, which intuitively requires completion of the tasks, but 
in the view of \filterOnGoing, where every task is ongoing, they can only result in removal of the marked tasks. 
An example of our new \pstate is shown in \cref{fig:a-table-involving-completion}. 
The $\PUT$ behavior of $\filterOnGoing$ converts $(\checkmark)$-marks to $(+)$-marks, 
reflecting the intention of the $(\checkmark)$ mark; recall that the $(+)$ mark indicates upsertion. 
The behavior of $\TaskInit$ and $\DUP$ is unchanged. 
For example, for the original source $\OSourceVar$ and the updated view in \cref{fig:a-table-involving-completion}, the $\PUT$ of $\TaskInit$ returns {the following tasks}, where the task with \texttt{001} is deleted and the one with \texttt{003} is completed. 
\fullversion{\[
\left(\text{\myfootnotesize\OrigTableAfterCompleteOG}\right)
\]}
Similarly, for the \VName{today} view, we consider the mark $(\Postpone)$, which intuitively means postponing; the mark does not convey the information of postponing days, as it is carried in the task itself. 
Similarly to the $(\checkmark)$ case, $(\Postpone)$-marked tuples specify that the marked tuples will not be presented in the view of $\filterToday$. 
Again, in the $\PUT$ execution, $\filterToday$ converts $(\Postpone)$ to $(+)$, requesting postponing. 

A caveat is that, when we compare such \pstates with \fstates, we cannot escape from the gap in expressiveness of user's intention between them. 
For example, on the view of \filterOnGoing, we cannot distinguish deletions and completions via \fstates as no completed tasks are allowed on the view. 
This means, on that view, $w \le t$ for a completion intention $w$ and a \fstate $t$ where the task to be completed in $w$ is not present. This is intrinsic: the difference between completion and deletion in the view of \filterOnGoing cannot be expressed in \fstates. 
However, this does not directly mean our lenses can garble such intentions. 
We prevent this by design: we allow \pstates also in $\GET$, which comes with other benefits~(\cref{sec:scenarios-partially-specified-states-in-get}).

\subsection{Partially-Specified States in \texorpdfstring{$\GET$}{GET}}
\label{sec:scenarios-partially-specified-states-in-get}

In \cref{sec:approach-overview}, we primarily focused on how $\PUT$ makes use of \pstates, as demonstrated in \cref{fig:running-example-our-put}. 
The $\GET$ direction becomes relevant when discussing compositional reasoning of well-behavedness, which concerns statements about the behavior of $\PUT$-then-$\GET$ and $\GET$-then-$\PUT$. 
Since the main purpose of \pstates is to improve $\PUT$'s behavior in handling multiple views, it is natural to ask why we care about them in $\GET$.
This design has both theoretical and practical benefits. 

Simplicity is one of the most apparent theoretical benefits. 
If $\GET$ operates only on \fstates and $\PUT$ propagates only \pstates, the original and updated states would belong to different domains, as $\GET \in S \to V$ and $\PUT \in S \times P \to Q$ similarly to update-update lenses~\cite{AhmanU17}, where $P$ and $Q$ are sets of \pstates corresponding to the sets $S$ and $V$ of \fstates, respectively. 
This, however, complicates well-behavedness (round-tripping) as we cannot pass the $\PUT$ results directly to $\GET$ and vice versa. We may need ways to connect $S$ to $P$ and $V$ to $Q$, and lenses must preserve such connections as well. 

This simplicity leads to a benefit that is both theoretical and practical: reusability. 
Our lenses follow classical lenses~\cite{FosterGMPS05,FGMPS07} computationally: in our framework, a lens between $S$ and $V$ is a pair of transformations $\GET \in S \to V$ and $\PUT \in S \times V \to S$ operating on the same domains (we shall ignore partiality of these transformations for now). The only difference is that we consider \pstates and partial ordering in $S$ and $V$ when reasoning about well-behavedness of lenses. 
Thanks to this design, we can reuse many lenses and lens combinators from the existing literature, though with the proof obligation of our version of well-behavedness. Lens composition is such an example. In a special case, a well-behaved lens in the classical lens framework~\cite{FosterGMPS05,FGMPS07} is also a well-behaved lens in our framework.
Another benefit is that we can reuse existing lens representations for our lenses. For example, we can tuple~\cite{Chin93,HuITT97} the transformations to bundle common computations for the source in $\GET$ and $\PUT$~\cite{Takeichi2009}. We might also use van Laarhoven lenses~\cite{OConnor11,vanLaarhovenLens} as used in the \texttt{lens} library in Haskell, or profunctor optics~\cite{PickeringGW17}.

The strength of the guarantee is another important benefit. If we restrict $\GET$ to work only on \fstates, the update preservation can only be stated by comparing \pstates with \fstates, and hence struggles to reject lenses that garble finer intentions that can only be represented in \pstates, such as one that replaces $(\checkmark)$-marked tuples with $(-)$-marked ones in $\PUT$. If $\GET$ of such a lens works also on \pstates, especially for those that the $\PUT$ returns, we can find that the $\PUT$ fails to preserve the user's intention by comparing the updated view and the view of the updated source. 

Another practical benefit of \pstates in $\GET$ is allowing competing updates on multiple views with certain precedence; 
in this case, our lenses will handle (delta-state~\cite{delta-CRDT}) CRDTs~\cite{CRDT}, which, roughly speaking, partially-ordered states among which $\vee$ is total (\ie, $\PUT$ of $\DUP$ never fails).

In exchange for these advantages, our framework requires programmers to define $\GET$ for partially-specified states. 
This would not be overly burdensome, 
as we define $\PUT \; (s,v)$ to return a partially-specified state with the knowledge that the information is sufficient for $\GET$ to return a meaningful result for $v$.
\section{Lenses for Partially-Specified States}
\label{sec:absence-aware-lenses}

In this section, as the main contribution of this paper, we describe our proposed lens framework, \emph{\our{} lenses} (\ourshort{}-lenses, for short), that can handle unspecified states like $\OViewOGVar$ and $\OViewDTVar$ in \cref{sec:scenarios}. %
The key point of our formalization is that it supports compositional reasoning, similarly to the original lens framework~\cite{FosterGMPS05,FGMPS07}.

\subsection{Preliminaries: Classical Lenses}
\label{sec:preliminaries}

We begin by reviewing the classical lens framework~\cite{FosterGMPS05,FGMPS07}, on which our framework is based. 
As mentioned in \cref{sec:scenarios-partially-specified-states-in-get}, the main difference lies in the laws and structures on domains.
In this paper, we consider partial functions between sets, rather than complete partial orders (CPOs) for simplicity and because $\PUT$ can fail for various reasons, such as conflicting updates on copies. 
Although this in effect rules out general recursions and allows only terminating functions, we believe that our discussions could be extended to CPOs straightforwardly.

A (partial) (asymmetric) \emph{classical lens} $\ell$ between a ``source'' set $S$ and a ``view'' set $V$ is a pair of partial functions $\GET \in S \to V$ and $\PUT \in (S \times V) \pto S$. 
Here, we used $\pto$ to emphasize that $\PUT$ can be partial, while we assume the totality of $\GET$ for simplicity. 
We write $\Lens{S}{V}$ for the set of all lenses  $\ell = (\GET, \PUT)$ between $S$ and $V$, and  $\GET \; \ell$ and $\PUT \; \ell$ for
their first and second components.
Intuitively,  $\GET \; \ell \in S \to V$ accesses a view from a source $s$; and $\PUT \;  \ell \in (S \times V) \pto S$
propagates a view update, which changes $\GET \; \ell \; s$ to some $v'$, into
a source update, which in turn changes $s$ to $\PUT \; \ell \; (s,v')$.
Since $\PUT$ may be partial, the update propagation may fail.

\newlength\MyMarginLawRA
\newlength\MyMarginLawRC
\newlength\MyMarginLawRS
\setlength{\MyMarginLawRA}{\widthof{(acceptability)}}%
\setlength{\MyMarginLawRC}{\widthof{(consistency)}}%
\setlength{\MyMarginLawRS}{\widthof{(stability)}}%
\newcommand{\InformalProp}[1]{\it{``#1''}}

Not all lenses are reasonable ones. The following laws establish the bidirectionality of lenses~\cite{FGMPS07,Hegner90,BancilhonS81}.
To provide a basis for our later discussions, we accompany these with informal properties (P) that they ensure, which we will aim to achieve for \our{} lenses. %
{\small\begin{gather*}%
  \forall s,s' \in S, \forall v' \in V.\;  \PUT \; \ell \; (s, v') = s' \; \Longrightarrow \; \GET \; \ell \; s' = v'\tag{consistency}\label{law:consistency}\\
  \text{\InformalProp{Updates to the source's view are preserved in the updated source.}}\hspace{\MyMarginLawRC}\hspace{-20pt}\tag{P1}\label{req:consistency}\\[\blanklineskip]
  \forall s \in S, \forall v \in V.\;  \GET \; \ell \;s = v \; \Longrightarrow \; \PUT \; \ell \; (s,v) = s\tag{acceptability}\label{law:acceptability}\\
  \text{\InformalProp{No updates to the source's view means no updates to the source.}}\hspace{\MyMarginLawRA}\hspace{-20pt}\tag{P2}\label{req:acceptability}
\end{gather*}}%
Intuitively, the \ref{law:consistency} property, also known as \textsc{PutGet}~\cite{FGMPS07}, states that the updated source $s'$ (that $\PUT$ returns) correctly reflects the updated view $v'$ (that $\PUT$ takes) in the sense that its view (that we $\GET$) matches with $v'$. 
The \ref{law:acceptability} property, also known as \textsc{GetPut}~\cite{FGMPS07}, states that, if there is no update on the view (that we $\GET$), there is no update (resulting from $\PUT$) on the source. 
The two round-tripping properties are collectively called \emph{well-behavedness}, and lenses satisfying them are called \emph{well-behaved}.
Sometimes, an additional law called \textsc{PutPut}~\cite{FGMPS07} is considered, which intuitively states that $\PUT$ preserves compositions of updates. We ignore the law, as it is generally considered too \fullversion{strong to permit useful transformations~\cite{FGMPS07,Keller87} in}{} state-based systems; we revisit this law in \cref{sec:related-work}\fullversion{, when we compare our proposed framework with operation-based systems}.
The \ref{law:acceptability} implies an important property, \ref{law:stability}, which is called \textsc{PutGetPut} in the literature~\cite{MuHT04aplas}. 
{\small\begin{gather*}
  \forall s_0,s \in S, \forall v \in V.\;  \PUT \; \ell \; (s_0, v) = s \; \Longrightarrow \; \PUT \; \ell \; (s, \GET \; \ell \; s) = s\tag{stability}\label{law:stability}\\
  \text{\InformalProp{One round-trip  leads to a stable source and view}} \hspace{\MyMarginLawRS}\tag{P3}  \label{req:stability}
\end{gather*}}%
When $\PUT \; \ell \; (s_0,v)$ succeeds, resulting in an updated source $s$, and then we extract its view with $\GET \; \ell \; s$, the resulting pair $(s, \GET \; \ell \; s)$ is in a \emph{stable state}: any further $\GET$ or $\PUT$ with them has no effect. In other words, one round-trip---\PUT then \GET---is enough to synchronize the two\fullversion{, and thus captures the entire update propagation process}.

\subsection{Domains: Partially-Ordered Sets with Identical Updates}
\label{sec:posets}

As mentioned in \cref{sec:introduction,sec:scenarios}, we use partially-ordered sets (posets, for short) for states, which are called \pstates and come with the notion of update preservation defined by the partial ordering $\le$. 
We also assume that our domains are equipped with a set of identical updates $\IDonSet{s}$ relative to $s$ in order to state our law(s) corresponding to \ref{law:acceptability}.
\begin{definition}\rm
A domain $P = (S, \le, \IDonName)$, which we call \emph{\IPoset}, is a triple where $(P, \le)$ is a poset and $\IDonName \subseteq S \times S$ is a reflexive subset of $(\le)$.
\end{definition}
\noindent 
For $P = (S, \le, \IDonName)$, we write $\Carrier{P}$ for $S$ and write $\IDonName_s$ for $\{ s' \mid (s', s) \in \IDonName \}$. The intuition behind the requirement $\IDonName \subseteq (\le)$ is that $s \le s'$, which means that $s$ is less specified than $s'$, can also be regarded as $s$ representing no update with respect to $s'$; conceptually, $\IDonName$ is a sub-partial-order of $\le$ that compares identical updates. We do not require the transitivity of $\IDonName$ just because we do not use the property in our formal development. 
\fullversion{(The antisymmetry of $\IDonName$ follows from $\IDonName \subseteq (\le)$.)}{}
The difference between the two relations matters when a \pstate represents more than a set of possible \fstates. 
When $P$ has the least element $\GO$ (or $P$ is lower-bounded), we also assume that $\GO \in \IDonSet{s}$ for any $s$, meaning that the update that specifies nothing can work as an identical update to any state. 

Simple examples of \IPosets include discrete posets where $(\le) = (=)$, which also implies $\IDonName = (=)$, and posets obtained by adding the distinct least element to a discrete poset, where the least element $\GO$ represents ``anything''. 
When \pstates just represent sets of possible \fstates, an \IPoset $(2^S, (\supseteq), (\supseteq))$ obtained from the power set of a given set $S$ of \fstates provides 
maximum flexibility. One may exclude $\emptyset$, which represents the ill-specification, from the state space in favor of failures of $\PUT$. The join $(\cap)$ becomes partial, then.
In these simple examples, $\IDonName$ is the same as $\le$, but the difference between the two becomes evident in 
non-trivial examples like below.

\begin{example}[Delimiting Identical Updates]
\label{ex:partial-pstate}
Often, a \pstate assumes a certain condition on its original (\fstateHead) state. In such a case, the difference between $\le$ and $\IDonName$ matters. 
To illustrate this, consider key-value mappings represented as partial functions in $K \pto V$ for a key set $K$ and a value set $V$. 
Consider \pstates that represent deletion requests of mappings, which are naturally represented by a subset $D$ of $K$. We can easily merge two deletion requests by the set union, and whether a deletion $D$ is reflected in $f$ is formalized in $D \cap \dom(f) = \emptyset$. 

There are some possibilities for this domain to permit both \fstate $f$ and \pstate $D$. 
A choice is $P_1 = ((K \pto V) \uplus 2^K, (\le), (\le))$, where $(\le)$ is the smallest reflexive relation that includes $(D,D')$ with $D \subseteq D'$ and $(D,f)$ with $D \cap \dom(f) = \emptyset$. 
However, since $P_1$ permits too many ``identical updates'', such as $K$ (which intuitively means ``delete everything'') for $f = \emptyset$, it is difficult or complicated to define lawful lenses, which must preserve identical updates. 
A more practical choice is $P_2 = ((K \pto V) \uplus 2^K, (\le), \IDonName)$, where $\IDonName$ is the smallest reflexive relation that includes $(D,D')$ with $D \subseteq D'$ and $(D,f)$ with $D = \emptyset$, ruling out deleting non-existing IDs in $\IDonSet{f}$. 
\qed 
\end{example}

We can even encode updates~\cite{DiskinXC11,Meertens98,AhmanU17,HofmannPW12} as \pstates by pairing them with their starting points. 
See \cref{sec:general-state-update-pairs} for the details.

Since there is a variety of domains as we have seen, our formalization is designed to be independent of any concrete representation of \pstates:
the only assumption on the domains (sets of source and view states) is that they form \IPosets.
 (In contrast, some concrete primitive lenses and combinators, such as those illustrated in \cref{sec:scenarios}, require concrete \IPoset or impose further assumptions to ensure functionality and/or well-behavedness.)
One might then wonder why we do not define their actions on \fstates like updates. We do, but instead of integrating this into the definition of our domains, we internalize it using certain primitive lenses called \initiators (\cref{sec:initiators}).

\subsection{(Lawless)  \OUR{} Lenses}
\label{sec:lawless-absence-aware-lenses}

Recall that the primary goal of this paper is to give a formal treatment of lenses that handle \pstateHead data.
Now that we have formalized \mbox{(possibly-)}\pstateHead data, we are ready to define (lawless-versions of) \our{} lenses.

\begin{definition}[(Lawless) \OUR{} Lenses]\label{def:lawless-absence-aware-lenses}\rm
For \IPosets $P$ and $Q$, a \emph{\our{} lens} (\ourshort{}-lens) $\ell$ between $P$ and $Q$ is a pair of functions
$\GET : \Carrier{P} \to \Carrier{Q}$ and $\PUT : \Carrier{P} \times \Carrier{Q} \pto \Carrier{P}$. \qed
\end{definition}
\noindent
We write $\MALens{P}{Q}$ for the set of \our{} lenses $\ell = (\GET, \PUT)$ between $P$ and $Q$, and $\GET \; \ell$ and $\PUT \; \ell$ for their first and second components. 
Definition \ref{def:lawless-absence-aware-lenses} suggests that, without laws, \our{} lenses are no different from
the classical lenses~\cite{FGMPS07}. %
In what follows, we sometimes simply call \ourshort{} lenses ``lenses'' unless confusion arises; they indeed are lawless lenses.

We show some examples of \our{} lenses (which indeed are lawful).
\begin{example}[Identity Lens]\label{ex:identity-lens}
Our simplest lens is the identity lens $\mathit{idL} \in \MALens{P}{P}$, defined as:
\begin{fullversionblock}
\begin{qedmath}
 \GET \; \mathit{idL} \; s = s \qquad
 \PUT \; \mathit{idL} \; (\dontcare, v) = v %
\end{qedmath}  
\end{fullversionblock}

\end{example}

\newcommand{\ConstL}[1]{\ensuremath{\var{constL}_{#1}}}

\begin{example}[Constant Lenses]
\label{ex:constant-lens}
Sometimes, a part of a view data is determined regardless of a source, such as element names in XML views. 
The constant lens $\ConstL{a} \in \MALens{P}{Q}$ for lower-bounded (\ie, having the least element $\GO$) with $a \in \Carrier{Q}$ is a useful building block for such a situation. 
\[
 \GET \; \ConstL{a} \; \dontcare = a \qquad
 \PUT \; \ConstL{a} \; (\dontcare, v) = \GO~\text{if}~\IDon{v}{a}
\]
Here, $\IDon{v}{a}$ instead of $v = a$ is intentional, and is required for the lens to be lawful; intuitively, this ensures that it works when composed with lenses that have a similar behavior to constant lenses. 
This lens is interesting for two reasons. First, its $\PUT$ is purposely partial (unless $\Carrier{Q} = \{a\}$). Second, by returning the least element $\GO$, it explicitly states that the source can be anything. \qed 
\end{example}

The next \our{} lens is more involved, and assumes an additional structure on an \IPoset for it to be well-behaved. 
\begin{definition}[Duplicable]\rm
We call an \IPoset $P$ \emph{duplicable} if it has a designated partial operator $\oplus \in \Carrier{P} \times\Carrier{P} \pto \Carrier{P}$ (called the \emph{merge operator} for $P$) satisfying both of the following conditions{.}
\begin{fullversionblock}
\begin{itemize}
  \item $x \oplus y = z$ implies $x \vee y = z$, and 
  \item for any $x$, $y$ and $z$, if $\IDon{x}{z}$ and $\IDon{y}{z}$, then $\IDon{x \oplus y}{z}$ holds. \qed
\end{itemize}  
\end{fullversionblock} 

\end{definition}
\noindent That is, $\oplus$ soundly computes joins, and must be total and closed (and thus must coincide with $\vee$) for identical updates to a fixed state. 
{The merge operator must be idempotent; since $\IDon{x}{x}$ holds for any $x$, $x \oplus x$ must be defined by the second condition and the result must be $x$ by the first condition.}{}
A sufficient condition of the duplicability is that $\oplus$ is complete ($(\oplus) = (\vee)$), and associative, in the sense that $x \oplus (y \oplus z)$ and $(x \oplus y) \oplus z$ coincide also in their definedness as well as their outcomes.
For simple \IPosets, this condition is easy to meet, but is often laborious or difficult to establish for tailored ones. 
The duplicability is preserved by the basic constructions of \IPosets, such as $P \x Q$, $P + Q$, and $P_\GO$ (the \IPoset obtained from $P$ by adding a new least element $\GO \not\in \Carrier{P}$), assuming point-wise operations for $\le$, $\IDonName$, and $\oplus$.

\begin{example}
Our duplication lens $\DUP \in \MALens{P}{(P \times P)}$ for duplicable $P$ is defined as:
\[
 \GET \; \DUP \; s = (s,s) \qquad
 \PUT \; \DUP \; \big(\dontcare, (s_1,s_2)\big) = s_1 \oplus s_2
\]
Here, the view data is simply a pair of source data. The $\GET$ is trivial, returning a view that contains two identical copies of the original source. For $\PUT$, we replace the source with the join of two (possibly) updated copies; the informal idea is that, by joining \pstates, their intentions, such as addition or deletion of particular tasks as seen in \cref{sec:approach-overview}, are merged into a unified one. 
For example with $P = \{1,2\}_\GO \x \{1,2\}_\GO = \{ (\Omega,\Omega), (1,\Omega), (\Omega, 2), (1, 2) \}$, with the point-wise ordering induced from $\GO \le 1,2$, meaning that 
$\GO$ is a ``no-op'' for the corresponding component, and with the merge operator $(\oplus) = (\vee)$ for which $\GO$ is the unit, 
we have $\PUT \; \DUP \; \big(s, (1, \Omega), (\Omega, 2)\big) = (1,\Omega) \oplus (\Omega, 2) = (1,2)$.
\qed
\label{ex:duplication-lens}
\end{example}

Finally, we can also define the \emph{composition} $\ell_1 \fatsemi \ell_2$  of two \our{} lenses $\ell_1 : \MALens{P}{Q}$ and $\ell_2 : \MALens{Q}{R}$ in the standard manner~\cite{FGMPS07} as below: %
\[\bb
 \GET \; (\ell_1 \fatsemi \ell_2) \; a = \GET \; \ell_2 \; (\GET \; \ell_1 \; a) \\
 \PUT \; (\ell_1 \fatsemi \ell_2) \; (a,c') = \PUT \; \ell_1 \; \big(a, \PUT \; \ell_2 \; (\GET \; \ell_1 \; a, c')\big)
\ee
\hspace{12pt}\bb
\begin{tikzcd}[nodes={inner sep=2pt},row sep=2pt, column sep=1cm]
{a} \arrow[r, "\GET \; \ell_1"] & {b} \arrow[r, , "\GET \; \ell_2"] & {c}\\
{a'} \putarrow["\PUT \; \ell_1"][7pt] & {b'} \putarrow["\PUT \; \ell_2"][7pt] & {c'}
\end{tikzcd}\ee
\]
The composition is associative and has the identity $\var{idL}$ in \cref{ex:identity-lens}.%
\begin{fullversionblock}
\footnote{When $\GET$ can be partial, the following condition is additionally required for $\var{idL}$ to be the right identity: $\PUT \; \ell \; (s,v) = s'$ implies $s \in \dom(\GET \; \ell)$. \fullversion{In other words, $\PUT$ is defined only for the original source $s$ for which $\GET$ is defined.}{} This does not pose any practical limitations as the invocation of $\PUT$ assumes that the successful execution of $\GET$.}
\end{fullversionblock}

\subsection{Laws: \OUR{} Consistency and Acceptability}
\label{sec:laws-absence-aware-acceptability-and-consistency}

Now, we focus on the laws for \our{} lenses. 
We first adapt 
\ref{law:consistency} and \ref{law:acceptability}~\cite{FGMPS07,BancilhonS81} of classical lenses
into the setting of \our{} lenses, while  keeping their direct informal intentions (\ref{req:consistency} and \ref{req:acceptability}).
We will leave \ref{law:stability} (\ref{req:stability}) for the moment; as we will see soon, \ref{req:consistency} and \ref{req:acceptability}'s extensions do not guarantee \ref{req:stability} in our setting, and we require an additional property.
\begin{fullversionblock}
Since domains are partially-ordered, it is natural to ask monotonicity of the $\GET$ and $\PUT$ transformations. While our versions of \ref{law:consistency} and \ref{law:acceptability} jointly enforce the monotonicity of $\GET$ (\cref{lemma:get-monotone}), we purposely allow non-monotone $\PUT$. 
\end{fullversionblock}

\subsubsection{\OUR{} Consistency}

We first focus on \ref{law:consistency}~(\ref{req:consistency})\fullversion{:}
\begin{fullversionblock}
{\small\begin{gather*}
  \forall s,s' \in S, \forall v' \in V.\:  \PUT \; \ell \; (s, v') = s' \; \Longrightarrow \; \GET \; \ell \; s' = v'\tag{consistency}\\
  \text{\InformalProp{Updates to the source's view are preserved in the updated source.}} \hspace{\MyMarginLawRC}\hspace{-20pt}\tag{P1}
\end{gather*}}\ignorespaces
\end{fullversionblock}
as it is directly related to the lens {\DUP} that plays a central role in our motivating example. 

Recall that $\DUP$ introduces the source sharing, which causes the violation of \ref{law:consistency} due to merging of updates on the copies. 
For example, we have $\PUT \; \DUP \; (s, ((1,\GO), (\GO,2))) = (1,2)$ for any $s$ and the view $((1,2), (1,2))$ of the updated source is different from 
either of the updated view $(1,\GO)$ and $(\GO, 2)$. 
As mentioned in \cref{sec:approach-overview}, we use $\le$ to state update preservation: $\PUT \; \ell \; (s,v') = s'$ must imply $v' \le \GET \; \ell \; s'$, meaning that the update intention in the updated view $v'$ is preserved in the view $\GET \; \ell \; s'$ of the updated source. 
To make the property compositional, we ensure $v' \le \GET \; \ell \; s''$ for any $s''$ with $s' \le s''$, ensuring that $\ell$ can be precomposed with lenses of the same kind. 
In summary, \our{} consistency for $\ell \in \MALens{P}{Q}$ is formalized as follows, satisfying~\ref{req:consistency}.
\begin{gather}
  \forall s,s' \in \Carrier{P}, \forall v' \in \Carrier{Q}.\;
  \PUT \; \ell \; (s, v') \le s'
  \;\Longrightarrow \;
  v' \le \GET \; \ell \; s'
  \tag{\ourshort{}-consistency}
  \label{law:aa-consistency}
 \end{gather}

\subsubsection{\OUR{} Acceptability}
Then, we adapt \ref{law:acceptability}~(\ref{req:acceptability})\fullversion{:} 
\begin{fullversionblock}
{\small\begin{gather*}
  \forall s \in S, \forall v \in V.\;  \GET \; \ell \;s = v \; \Longrightarrow \; \PUT \; \ell \; (s,v) = s\tag{acceptability}\\
  \text{\InformalProp{No updates to the source's view means no updates to the source.}} \hspace{\MyMarginLawRA}\hspace{-20pt}\tag{P2}
\end{gather*}}\ignorespaces
\end{fullversionblock}
It is true that the original version is compositional and leads to the stability as requested, but it is too strong for our purpose for the following two reasons. 
\begin{itemize}
 \item We prefer $\PUT$ to return the smallest possible results, so that further merging by $\vee$ is more likely to succeed. 
 (This is not true for a general poset but typically holds for our domains.) An extreme example is $\ConstL{42}$ in \cref{ex:constant-lens}, where 
 $\PUT \; \ConstL{42} \; (s_0, 42) = \GO$, saying that the lens does not contain the updated source at all and 
 any merge with $\GO$ always succeeds as $\GO \vee s = s \vee \GO = s$. 
 \item The original law hampers natural behavior of a lens as an update translator. Recall that, the $\PUT$ of the filter lens $\filterOnGoing$ in \cref{sec:approach-overview} passes insertion requests, \ie, $(+)$-marked tuples, on the view into the source intact. This seemingly innocuous behavior violates the original law, because the original source may contain update requests which are not representable on the view, such as insertion requests of completed tasks. Its $\GET$ or $\PUT$ must filter out or reject such updates for update preservation.
\end{itemize}

Thus, we adapt the original \ref{law:acceptability} law to these behaviors. Following the informal requirement \ref{req:acceptability}, which says that no-update, or identical updates, are preserved by $\PUT$, we give \our{} acceptability as follows. 
\[
  \forall s \in \Carrier{P}, \forall v \in \Carrier{Q}.\;
  \IDon{v}{\GET \; \ell \; s}
  \; \Longrightarrow \;
  \IDon{\PUT \; \ell \; (s, v)}{s}
 \tag{\ourshort{}-acceptability}
 \label{law:aa-acceptability}
\]
The original version corresponds to the case where $\IDonName$ is equality ($=$), or  equivalently $\IDonSet{s} = \{s\}$ for any $s$, on both domains.

\subsubsection{\OUR{} Weak Well-Behavedness}

So far, we have obtained \our{} versions of \ref{law:consistency} and \ref{law:acceptability},
which are the standard round-tripping properties used in this literature~\cite{BancilhonS81,Hegner90,FGMPS07}.
We collectively call the two properties \emph{weak well-behavedness}. We call it ``weak''
because it does not cover \ref{law:stability}, which we will recover in \cref{sec:absence-aware-stability} by adding a law.
Although it is ``weak'', when $(\le)$ is $(=)$, which implies $\IDonName$ is also $(=)$ as it is a reflexive subset of $(\le)$, the weak well-behavedness coincides with the original well-behavedness. In this sense, the weak well-behavedness is a clean generalization of the original.

We state {some}{} properties of weakly well-behaved \ourshort{} lenses.
\fullversion{{First}, their two defining properties are closed under the lens composition:}
\begin{lemma}
Both \ref{law:aa-acceptability} and \ref{law:aa-consistency} are closed under the lens composition. 
\label{lemma:acceptability-and-consistency-are-compositional}
\end{lemma}%
\begin{fullversionblock}
\begin{proof}[Sketch]
See the diagrams below; the black parts are given, and the red will be shown.
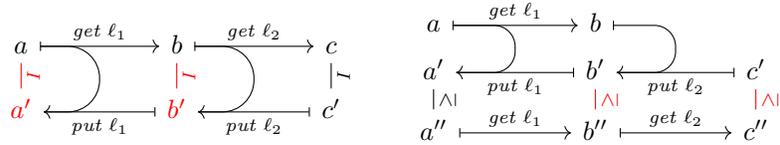
\[
  \begin{tikzcd}[row sep=10pt, column sep=1.5cm]
    {a}
    \arrow[r, mapsto, "\GET \; \ell_1"]
    \arrow[from=d, red, no head, "\IDonName"', sloped]
    &
    {b}
    \arrow[r, mapsto, "\GET \; \ell_2"]
    \arrow[from=d, red, no head, "\IDonName"', sloped]
    &
    {c}
    \arrow[from=d, no head, "\IDonName"', sloped]
    \\
    |[red]| {a'}
    \arrow[from=r, mapsto, "\PUT \; \ell_1"]
    \arrow[from=r, phantom, ""{coordinate, name=AppBpp}]
    \arrow[from=u, no head, rounded corners=12pt, to path={(\tikztostart.east) -| (AppBpp)\tikztonodes -- (\tikztotarget)}]
    &
    |[red]| {b'}
    \arrow[from=r, mapsto, "\PUT \; \ell_2"]
    \arrow[from=r, phantom, ""{coordinate, name=AppBpp}]
    \arrow[from=u, no head, rounded corners=12pt, to path={(\tikztostart.east) -| (AppBpp)\tikztonodes -- (\tikztotarget)}]
    &
    {c'}
  \end{tikzcd}
  \qquad
  \begin{tikzcd}[row sep=8pt, column sep=1.5cm]
    {a}
    \arrow[r, mapsto, "\GET \; \ell_1"]
    &
    {b}
    &
    {}
    \\[-5pt]
    {a'}
    \arrow[from=r, mapsto, "\PUT \; \ell_1"]
    \arrow[from=r, phantom, ""{coordinate, name=AppBpp}]
    \arrow[from=u, no head, rounded corners=8pt, to path={(\tikztostart.east) -| (AppBpp)\tikztonodes -- (\tikztotarget)}]
    \arrow[from=d, no head, "\ge"', sloped]
    &
    {b'}
    \arrow[from=r, mapsto, "\PUT \; \ell_2"]
    \arrow[from=r, phantom, ""{coordinate, name=AppBpp}]
    \arrow[from=u, mapsto, no head, rounded corners=8pt, to path={(\tikztostart.east) -| (AppBpp)\tikztonodes -- (\tikztotarget)}]
    \arrow[from=d, red, no head, "\ge"', sloped]
    &
    {c'}
    \arrow[from=d, red, no head, "\ge"', sloped]
    \\
    {a''}
    \arrow[r, mapsto, "\GET \; \ell_1"]
    &
    {b''}
    \arrow[r, mapsto, "\GET \; \ell_2"]
    &
    {c''}
  \end{tikzcd}
\]
To show the \ref{law:aa-acceptability} of $\ell_1 \fatsemi \ell_2$, we first use the \ref{law:aa-acceptability} of $\ell_2$ to show $\IDon{b'}{b}$, and then use the \ref{law:aa-acceptability} of $\ell_1$ to conclude $\IDon{a'}{a}$.
To show the \ref{law:aa-consistency} of $\ell_1 \fatsemi \ell_2$, we
first use the \ref{law:aa-consistency} of $\ell_1$ and then use that of $\ell_2$ to show $b' \le b''$ and $c' \le c''$ in turn.
\end{proof}

{Second}, $\GET \; \ell$ must be monotone for $\ell$ to be weakly well-behaved.
\begin{lemma}%
For a weakly well-behaved \ourshort{}-lens $\ell$, if $\GET \; \ell \; s = v$
and $s \le s'$, we have $v \le \GET \; \ell \; s'$. \qed
\label{lemma:get-monotone}
\end{lemma}
\begin{corollary}
For a weakly well-behaved \ourshort{}-lens $\ell$, $\GET \; \ell$ is monotone. \qed
\end{corollary}  
\end{fullversionblock}

\newcommand{\InitO}{\ensuremath{\var{init}_\GO}\xspace}

\begin{fullversionblock}  
In contrast, our $\PUT$ is not necessarily monotone in the second argument. 
Typically, \initiators violate the condition as they often try to keep the original source information as much as possible; thus, when they take a smaller \pstate, more information of the original source remains in the updated source. 
For example, for $\TaskInit$ in \cref{sec:scenarios}, 
observe that $\PUT \; \TaskInit \; (\OSourceVar, \GO) = \OSourceVar$ while $\PUT \; \TaskInit \; (\OSourceVar, w_\mathrm{merged}) = \OSourceVar''$ with $\OSourceVar \not\le \OSourceVar''$. 
{%
There is also a lawful lens whose $\PUT$ is not monotone in the first argument (see \cref{sec:non-monotone-in-first-arg});}{}
unlike the second-argument case, we have found no practical reason to have lenses of this kind---they are allowed just because our requirements (laws) are minimal.
\end{fullversionblock}

\begin{fullversionblock}
{Third}, repeated round-tripping does not change the view. This is called \textsc{GetPutGet} in the literature~\cite{MuHT04aplas}, and is implied by \ref{law:acceptability} (\textsc{GetPut}) in the classical setting.
\begin{lemma}[Stability in View]
For a weakly well-behaved \our{} lens $\ell$, for any $s$ with $\GET \; \ell \; s = v$, we have $\GET \; \ell \; (\PUT \; \ell \; (s, v)) = v$. \qed
\label{lemma:view-stability}
\end{lemma}

{Fourth}, similarly to the classical lenses~\cite{FischerHP15}, $\PUT$ uniquely determines $\GET$, assuming weak well-behavedness.
\begin{lemma}
  Let $\ell \in \MALens{P}{Q}$ be a weakly well-behaved \our{} lens, and
  $V_s$ be the set $\{ v \in |Q| \mid \exists s_0, s_1 \in |P|.\, \PUT \; \ell \; (s_0, v) = s_1 \wedge s_1 \le s \}$.
  Then, we have $\GET \; \ell \; s = \max V_s$. 
  \label{lemma:put-determines-get}
\end{lemma}
\Cref{lemma:put-determines-get} provides a guideline to define $\GET$ for \pstates when the corresponding $\PUT$ is given; $\GET$ is in fact determined by $\PUT$, and we have no choice. 
While computing $V_s$ is often not straightforward due to overshooting ($s_1 \le s$), 
the lemma helps filter out candidates; for example, if $\PUT$ returns the same $s$ for different $v_1$ and $v_2$ with $v_1 \le v_2$, its $\GET$ cannot return $v_1$ as $v_1$ cannot be $\max V_s$. 
In contrast to $\PUT$, $\GET$ does not always determine $\PUT$, also similarly to the classical lenses. As we will see soon, well-behaved classical lenses are also well-behaved (and hence weakly well-behaved) \our{} lenses (\cref{prop:classical-well-behaved}). 
\end{fullversionblock}

\subsection{Additional Law: \OUR{} Stability}
\label{sec:absence-aware-stability}

Now we have laws \ref{law:aa-acceptability} and \ref{law:aa-consistency} that satisfy properties \ref{req:acceptability} and \ref{req:consistency}, respectively. Our next goal is to find a law that adapts \ref{law:stability} to achieve \ref{req:stability}, while also supporting compositional~reasoning. 
\begin{fullversionblock}
\begin{gather*}
  \forall s_0, s \in S, \forall v \in V.\;  \PUT \; \ell \; (s_0, v) = s \; \Longrightarrow \; \PUT \; \ell \; (s, \GET \; \ell \; s) = s \tag{stability}\\
  \text{\InformalProp{One round-trip leads to a stable source and view}} \hspace{\MyMarginLawRS}\tag{P3}
\end{gather*}  
\end{fullversionblock}
\fullversion{This}{} stability law is important. It says that one round-trip---updating the source with \PUT, and then acquiring its view with \GET---captures the entire update process needed to reach a stable state $(s, \GET \; \ell \; s)$: any further \GET or \PUT on the pair has no effect.

\subsubsection{Issue: Non-Stability}\label{sec:non-stability}

As noted earlier, weak well-behavedness alone does not guarantee stability. 
The \ref{law:acceptability} property implies \ref{law:stability}, but, in our case, we only have a weaker form of \ref{law:acceptability}, which makes stability an issue.
The challenge for \our{} lenses is that, $\PUT$ may result in a \pstate, like with the lens \ConstL{42}. 
By using \ref{law:aa-acceptability}, we have $\IDon{\PUT \; \ell \; (s, \GET \; \ell \; s)}{s}$, which implies $\PUT \; \ell \; (s, \GET \; \ell \; s) \le s$, but it can happen that $\PUT \; \ell \; (s, \GET \; \ell \; s) < s$. %
Although $\ConstL{42}$ happens to be stable,\fullversion{ as the \ref{law:stability} concerns the case where the original source is a result of $\PUT$ (thus, $s = \GO$),} the underlying behavior---that $\PUT$ can return a smaller state than its original source---can be exploited to make unstable lenses, as illustrated below. %

\begin{example}[Unstable Lens]
  \label{ex:non-stabilizing-lens}
Let \(\UnitType\) be the singleton \IPoset 
whose only element is $(\,)$ and $N$ be the set $\{\underline{0},\underline{1},\underline{2},\dots\}$ where $\underline{i} \le \underline{j}$ if and only if $i \le j$, and $\IDonName = (\le)$.\footnote{Hence, $N$ is the set of natural numbers with the standard ordering. We do not use $\mathbb{N}$ here, as $1$ and $2$ in $\mathbb{N}$ are incompatible in specifiedness in the usual sense{; they are both \fstates.}} Intuitively, an element $\underline{n} \in N$ represents a precedence of updates (without payload). 
Consider a lens $\var{bad} : \MALens{N}{\UnitType}$ defined as:
\[
  \GET \; \var{bad} \; \dontcare = (\,) \qquad
  \PUT \; \var{bad} \; (s,(\,)) = \begin{cases}
    \underline{0} & \text{if}~s = \underline{0} \\
    \underline{k} & \text{if}~s = \underline{k+1}
   \end{cases}
\]
Since $\PUT \; \var{bad} \; (\underline{n}, (\,))$ with non-zero $n$ results in $\underline{n-1}$, 
an $n$-fold application of $\PUT \; \var{bad} \; (-,(\,))$ is needed for $\underline{n}$ to reach a fixed point $\underline{0}$, failing to fulfill \ref{req:stability}.

Regardless, the lens $\var{bad}$ still satisfies both \ref{law:aa-acceptability} and \ref{law:aa-consistency};
the former reduces to $\IDon{\PUT \; \var{bad} \; (s, (\,))}{s}$, which is obvious, and the latter is trivial as the only possible view is the unit value $(\,)$. Hence, weak well-behavedness is not enough to rule out this undesirable lens.
\qed
\end{example}

\begin{myremark}%
If the final source is discrete---having only concrete states, which is often desirable---\ref{law:stability} already holds. This is because  $(\le) = \IDonName = (=)$ in a discrete domain, and thus \ref{law:aa-acceptability} ensures $\PUT \; \ell \; (s', \GET \; \ell \; s') = s'$.
However, the $\PUT$ of an unstable lens may not always provide a ``definitive answer'', demonstrated by \cref{ex:non-stabilizing-lens} where performing multiple round-trips can gradually refine the source. Thus, for the composition of unstable lenses,
while the overall \PUT could fail, extra local round-trips might lead to success in propagating updates.
We also note that, as mentioned in \cref{sec:scenarios-partially-specified-states-in-get}, having \spstates in the final source is useful when a lens is supposed to accept any simultaneous updates on multiple views by using CRDTs~\cite{CRDT,delta-CRDT}.
\end{myremark}

\subsubsection{\OUR{} Stability}

Thus, we need an additional property for \ref{req:stability}.
The issue with directly enforcing the \ref{law:stability} law is that it conflicts with compositional reasoning, a core principle underlying lenses as an approach to bidirectional transformations.
It is known that \ref{law:stability} by itself is not closed under composition~(see Matsuda \etal~\cite[Appendix B]{MatsudaW15}).

Since \ref{law:aa-acceptability} already ensures $\IDon{\PUT \; \ell \; (s', \GET \; \ell \; s')}{s'}$, which implies $\PUT \; \ell \; (s', \GET \; \ell \; s') \le s'$, 
it suffices to enforce $s' \le \PUT \; \ell \; (s', \GET \; \ell \; s')$ to entail \ref{law:stability}. 
To enforce this compositionally, assuming \ref{law:aa-consistency} and \ref{law:aa-acceptability}, 
we require the following law. 
\begin{gather*}
  \begin{split}
  &\forall s_0,s, s',s'' \in \Carrier{P}.\, \forall v, v'' \in \Carrier{Q}.\:\\
  &\quad
  \big(\left(\PUT \; \ell \; (s_0, v) = s \le s'\right)
  \; \wedge \;
  \left(v \le v'' \in \IDonSet{\GET \; \ell \; s'}\right)
  \; \wedge \;
  \left(\PUT \; \ell \; (s',v'') = s''\right)
  \big) \\
  &\quad {\Longrightarrow{}} \;
  s \le s''
  \end{split}
  \tag{\ourshort{}-stability}
  \label{law:aa-stability}
\end{gather*}
\begin{wrapfigure}[5]{r}{3.4cm}
\ifFullVersion
\ifdim\pagetotal=0pt \else\vspace{-\intextsep}\fi
\else 
\fi 
\(
\begin{tikzcd}[nodes={inner sep=2pt},row sep=10pt, column sep=1.1cm]
  {s'}
  \arrow[r, mapsto]
  &
  {v'} %
  \arrow[blue, dd, no head, bend left=60, "\le"', sloped]
  \\
  {\textcolor{blue}{\smash{{}^\exists}} s''}
  \arrow[u, blue, no head, "\IDonName"', sloped]
  \arrow[from=r, mapsto]
  \putarrow
  &
  {v''}
  \arrow[u, no head, "\IDonName"', sloped]
  \\
  {s}
  \arrow[u, red, no head, "\le"', sloped]
  \arrow[uu, no head, bend left=60, "\le", allow upside down, sloped]
  \arrow[from=r, phantom, ""{coordinate, name=AB}]
  \arrow[from=r, densely dotted,  rounded corners=10pt, to path={ ([yshift=2ex]AB) .. controls (AB) .. (\tikztotarget) }]
  \arrow[from=r, mapsto]
& {v}
  \arrow[u, no head, "\le"', sloped]
\end{tikzcd}
\)
\end{wrapfigure}
This law may be easier to understand with the diagram on the right, where the concluded parts are colored red, and the parts derived from \ref{law:aa-acceptability} and \ref{law:aa-consistency} are colored blue. Here, we write $s$ for $\PUT \; \ell \; (s_0, v)$ and $v'$ for $\GET \; \ell \; s'$. 
We place greater elements in upper rows to highlight ordering among elements. (We omit $s_0$ as it is irrelevant for the ordering, and denote the omitted inputs by dotted lines in the diagram.)

The law is designed to be preserved under composition. 
The conditions $s \le s'$ and $\IDon{v''}{\GET \; \ell \; s'}$ (rather than $s=s'$ and $v'' = \GET \; \ell \; s'$) mirror the designs of \ref{law:aa-consistency} and \ref{law:aa-acceptability}. 
In contrast, the condition $v \le v''$, which is enforced by the composition, 
is crucial to make the law practical; otherwise, even $\var{idL} \in \MALens{P}{P}$ in \cref{ex:identity-lens} does not satisfy the property for general $P$ (\eg, take $s_0 = \GO$, $v = 1$, $s' = s = 1$, and $v'' = \GO$ to get $s'' = \GO$).
The definedness of $\PUT \; \ell \; (s',v'')$ appears as a requirement instead of a conclusion of the law, which does not make a difference as it is implied by \ref{law:aa-acceptability}.

\fullversion{Formally, we have \cref{lemma:composition-preserves-rput-get-put,lemma:aa-stability-implies-stability} below.}
\begin{lemma}
For $\ell_1$ and $\ell_2$ that satisfies all of \ref{law:aa-acceptability}, \ref{law:aa-consistency}, and \ref{law:aa-stability}, $\ell_1 \fatsemi \ell_2$ also satisfies \ref{law:aa-stability}.
\label{lemma:composition-preserves-rput-get-put}
\end{lemma}

\begin{myproof}
{\raggedright Consider the diagram on the right.\par}
\begin{wrapfigure}[5]{r}{5.3cm}
\ifdim\pagetotal=0pt \else\vspace{-1.2\intextsep}\fi
\vspace{-0.5\baselineskip}
\(
  \begin{tikzcd}[row sep=7.5pt, column sep=1.5cm, nodes={inner sep=1.5pt}]
    {a'}
    \arrow[r, mapsto, "\GET \; \ell_1"]
    &
    {b'}
    \arrow[r, mapsto, "\GET \; \ell_2"]
    &
    {c'}
    \\
    {a''}
    \putarrow["\PUT \; \ell_1"]
    &
    {b''}
    \arrow[u, line width=1pt, red, "I"', sloped, no head]
    \putarrow["\PUT \; \ell_2"]
    &
    {c''}
    \arrow[u, no head, "I"', sloped]
    \\
    {a}
    \arrow[uu, no head, "\le", sloped, bend left=60, allow upside down]
    \arrow[from=r, phantom, ""{coordinate, name=AB}]
    \arrow[from=r, densely dotted,  rounded corners=8pt, to path={ ([yshift=1.5ex]AB) .. controls (AB) .. (\tikztotarget) }]
    \arrow[from=r, mapsto, "\PUT \; \ell_1"]
    \arrow[u, line width=1pt, red, "\le"', sloped, no head]
    &
    {b}
    \arrow[u, line width=1pt, red, "\le"', sloped, no head]
    \arrow[uu, thick, red, no head, bend left=60, "\le", pos=0.7, sloped, allow upside down, crossing over]
    \arrow[from=r, phantom, ""{coordinate, name=AB}]
    \arrow[from=r, densely dotted,  rounded corners=8pt, to path={ ([yshift=1.5ex]AB) .. controls (AB) .. (\tikztotarget) }]
    \arrow[from=r, mapsto, "\PUT \; \ell_2"]
  & {c}
    \arrow[u, "\le"', sloped, no head]
  \end{tikzcd}
\)
\end{wrapfigure}
\noindent This diagram illustrates the execution of $\PUT \; (\ell_1 \fatsemi \ell_2) \; (a_0, c) = a$, $\GET \; (\ell_1 \fatsemi \ell_2) \; a' = c'$, and $\PUT \; (\ell_1 \fatsemi \ell_2) \; (a',c'') = a''$, with $a \le a'$ and $c \le c'' \le c'$. 
Here, we use thick red arrows for the parts to be concluded.

We have $b \le b'$ by the \ref{law:aa-consistency} of $\ell_1$ and $a \le a'$. Then,
we use the \ref{law:aa-stability} of $\ell_2$ to obtain $b \le b''$.
We also have $\IDon{b''}{b'}$ by the \ref{law:aa-acceptability} of $\ell_2$ and $c'' \le c'$.
Then, we have $a \le a''$, as required, by the \ref{law:aa-stability} of $\ell_1$.
\end{myproof}

\begin{lemma}
If a \our{} lens $\ell$ satisfies all of \ref{law:aa-acceptability}, \ref{law:aa-consistency}, and \ref{law:aa-stability}, it also satisfies \ref{law:stability}. 
\label{lemma:aa-stability-implies-stability}
\end{lemma}

\begin{fullversionblock}
\begin{proof}
Suppose $s = \PUT \; \ell \; (s_0,v)$ for some $s_0$ and $v$. Then, we prove that $\PUT \; \ell \; (s, \GET \; \ell \; s)$ is defined to be $s$. 
Then, \ref{law:aa-consistency} ensures $v\le \GET \; \ell \; s$ and \ref{law:aa-acceptability} ensures the definedness of $s'' = \PUT \; \ell \; (s, \GET \; \ell \; s)$ together with $\IDon{s''}{s}$. Now, we are ready to use \ref{law:aa-stability} with $s' = s$ and $v'' = \GET \; \ell \; s$ to obtain $s \le s''$. 
Since we have $s'' \le s$ by $\IDon{s''}{s}$, we have $s = s'' = \PUT \; \ell \; (s, \GET \; \ell \; s)$ by the antisymmetry of $\le$. \qed 
\end{proof}  
\end{fullversionblock}

Now, we are ready to define well-behavedness of \our{} lenses.
\begin{definition}[Well-Behavedness of \Our{} Lenses]\rm
A \our{} lens $\ell$ is well-behaved if it is weakly well-behaved and satisfies \ref{law:aa-stability}. \qed
\end{definition}
\noindent
In other words, a well-behaved \our{}  lens satisfies \ref{law:aa-acceptability}, \ref{law:aa-consistency} and \ref{law:aa-stability}. 
\fullversion{These properties describe the behavior of a lens in the situations of \GET-then-\PUT, \PUT-then-\GET, and \PUT-then-\GET-then-\PUT, respectively.}

\begin{theorem}
The well-behavedness is closed under composition. That is,
$\ell_1 \fatsemi \ell_2$ is well-behaved when $\ell_2$ and $\ell_1$ are. \qed 
\end{theorem}

When $P$ of $\ell \in \MALens{P}{Q}$ is discrete (\ie, $(\le) = \IDonName = (=)$), \ref{law:aa-stability} is implied by \ref{law:aa-acceptability}; in this case, 
well-behavedness and weak well-behavedness coincide. 
When $Q$ is discrete in addition, our (weak) well-behavedness 
coincides with the classical well-behavedness, showing that our formalization is a \emph{clean extension} of the classical lens framework.
\begin{lemma}
For $\ell \in \MALens{P}{Q}$ with discrete $P$,
$\ell$ is weakly well-behaved if and only if it is well-behaved. \qed
\label{prop:classical-well-behaved}
\end{lemma}
\begin{lemma}
A classical lens $\ell \in \Lens{A}{B}$ is well-behaved if and only if $\ell \in \MALens{(A,=,=)}{(B,=,=)}$ is well-behaved as a \ourshort-lens. \qed
\end{lemma}

\subsection{Examples of Well-Behaved \OUR{} Lenses}

Then, we show some examples of the well-behaved \our{} lenses.

\begin{example}[Identity Lens]
The identity lens $\var{idL} \in \MALens{P}{P}$ in \cref{ex:identity-lens} is well-behaved for any \IPoset $P$. \qed
\end{example}

\begin{example}[Constant Lens]
The constant lens $\ConstL{a} \in \MALens{P}{Q}$ for lower-bounded $P$ in \cref{ex:constant-lens} is well-behaved. 
\fullversion{The \ref{law:aa-acceptability} is straightforward as its $\PUT$ is defined for any $\IDon{v}{a}$ and returns the least element. Also, \ref{law:aa-consistency} is straightforward as its $\GET$ always returns $a$ regardless of the input. Since $\PUT \; \ell \; (s_0, v) = \Omega$ for any $s_0$ and $v$, \ref{law:aa-stability} is also straightforward: we have $s \le s''$ regardless of $s''$.} \qed 
\end{example}

\begin{example}[Duplication Lens]
The duplication lens $\delta \in \MALens{P}{(P \times P)}$ in \cref{ex:duplication-lens} is well-behaved if $P$ is duplicable.

\begin{fullversionblock}  
To ensure \ref{law:aa-acceptability}, it suffices to check if $\IDon{\PUT \; \DUP \; (s, (v_1,v_2)) = (v_1 \oplus v_2)}{s}$ holds for any $(v_1,v_2)$ with $\IDon{v_i}{s}$ ($i = 1,2$). 
Since $P$ is duplicable, $v_1 \oplus v_2$ is defined and satisfies $\IDon{(v_1 \oplus v_2)}{s}$. 
To ensure \ref{law:aa-consistency}, since $\GET$ is monotone, it suffices to check if $(v_1,v_2) \le \GET \; \DUP \; s = (s,s)$ holds when $s = \PUT \; \DUP \; (s_0, (v_1, v_2)) = (v_1\oplus v_2, v_1 \oplus v_2)$ is defined (and thus we can replace $\oplus$ with $\vee$). This follows immediately from the definition of $\vee$. 
To ensure \ref{law:aa-stability}, it suffices to show $(v_1 \oplus v_2) \le (v'_1 \oplus v'_2)$ with $v_i \le v_i'$ ($i = 1,2$) when both sides of $\le$ are defined (and thus we can replace $\oplus$ with $\vee$). Since $v_i \le v'_i \le (v'_1 \vee v'_2)$ holds for $i = 1,2$, we have $(v_1 \vee v_2) \le (v'_1 \vee v'_2)$ as $v_1 \vee v_2$ is the least upper bound of $v_1$ and $v_2$.
\end{fullversionblock}
\qed
\end{example}

\begin{nonexample}
The \fullversion{undesirable} lens $\var{bad}$ in \cref{ex:non-stabilizing-lens} is not well-behaved as it violates \ref{law:aa-stability}.
Take $s_0 = \underline{2}$ for which $s = \PUT \; \var{bad} \; (s_0, (\,)) = \underline{1}$, and take $s' = s$, then we have $\PUT \; \var{bad} \; (s', (\,)) = \PUT \; \var{bad} \; (\underline{1}, (\,)) = \underline{0} \not\ge  \underline{1} = s = s'$, violating the property. \qed
\end{nonexample}

\begin{example}[Untagging]
Many primitives lenses that we so far have seen do not use their first arguments in $\PUT$. 
Of course, this is not always the case. The following lens $\UnTagS \in \MALens{(P + P)}{P}$ is such an example.
\[
\bb 
  \GET \; \UnTagS \; (\InL \; s_1) = s_1 \\ 
  \GET \; \UnTagS \; (\InR \; s_2) = s_2 \\
\ee
\qquad 
\bb 
 \PUT \; \UnTagS \; (\InL \; \dontcare) \; s = \InL \; s \\
 \PUT \; \UnTagS \; (\InR \; \dontcare) \; s = \InR \; s 
\ee
\]
Observe that its $\PUT$ determines the tag only by using the source information; this is why the lens is called $\UnTagS$.

\begin{fullversionblock}
We also have a general version $\UnTag \; \phi_1 \; \phi_2 \in \MALens{(P_{\phi_1} + P_{\phi_2})}{P}$
that uses both source and view information to determine the tag with monotone predicates $\phi_1$ and $\phi_2$. 
{\small\[
 \bb  
   \PUT \; (\UnTag \; \phi_1 \; \phi_2) \; s \; v = \begin{cases}
    \InL \; v & \text{if}~(s = \InL \; \dontcare \wedge \phi_1 \; v) \vee (s = \InR \; \dontcare \wedge \phi_1 v \wedge \neg (\phi_2 \; v)) \\
    \InR \; v & \text{if}~(s = \InR \; \dontcare \wedge \phi_2 \; v) \vee (s = \InL \; \dontcare \wedge \phi_2 v \wedge \neg (\phi_1 \; v))
  \end{cases}
 \ee 
 \]}%
Here, $P_\phi$ denotes the \IPoset whose elements are $\{ x \in \Carrier{P} \mid \phi \; x\}$.
They all are well-behaved for any \IPoset $P$.  
\end{fullversionblock}

The lens {\UnTagS}\fullversion{/{\UnTag}}{} serves as a building block of conditional branching. 
Due to space limitation, we omit the concrete definitions, but the idea is:  
we examine a condition by using a lens $\MALens{S}{(A + B)}$, perform computation on each case by lenses $\MALens{A}{C}$ and $\MALens{B}{C}$ to obtain $C + C$, 
and then finally erase the tag by $\UnTagS$\fullversion{/$\UnTag$ (in case of $\UnTag$, $C$ may be refined by different predicates for branches)}. This behavior corresponds to $\var{ccond}$\fullversion{/$\var{cond}$}{} in the classical lens~\cite{FGMPS07}. \qed 
\end{example}

\subsection{Relating Partial-Ordering and Updates: \INITIATORs}
\label{sec:initiators}

We give the formal definition of \initiators. Although they are no different from other lenses in our framework, their general form illustrates the relationship among \pstates, \fstates, updates, and partial orders. 

As mentioned informally in \cref{sec:approach-overview}, the definitions of \initiators $\var{init} \in \MALens{S}{P}$ with $S \subseteq P$ for a discrete $S$ share the same pattern\fullversion{. 
\[
\bbt
\GET \; \var{init} \; s     &= s \qquad\qquad  
\PUT \; \var{init} \; (s,v) &= \ApplyUpd{v}{s}
\ee\]}%
{}
Namely, its $\GET$ embeds \fstates $S$ in (possibly) \pstates $P$, and its $\PUT$ implements the semantics of \pstates as updates via $(\ApplyUpd{}{}) \in P \to S \pto S$.
Since $S$ is discrete, for $\var{init}$ to be well-behaved, it suffices to satisfy \ref{law:aa-acceptability} and \ref{law:aa-consistency}, which reduce to the following two properties: 
\begin{align*}
& \forall v \in \Carrier{P}, s \in \Carrier{S}.\; \IDon{v}{s} \Longrightarrow \ApplyUpd{v}{s} = s 
\tag{u-acceptability}\label{law:u-acceptability}\\
& \forall v \in \Carrier{P}, s,s' \in \Carrier{S}.\; \ApplyUpd{v}{s} = s' \Longrightarrow v \le s' 
\tag{u-consistency}\label{law:u-consistency}
\end{align*}
Intuitively, $\IDon{v}{s}$ means that $v$ represents no update with respect to $s$---this is what \ref{law:u-acceptability} ensures. 
The \ref{law:u-consistency} property states that the application $\ApplyUpd{v}{s}$ of an update $v$ to a state $s$ must, if it succeeds, preserve the intention given as $v$.

\section{Scenario Revisited: Formal Implementation}
\label{sec:revisiting}
\newcommand{\KVTree}{\con{Tr}}
\newcommand{\KVTreeSpec}{\con{TrSpec}}

We are now ready to revisit the example from \cref{sec:scenarios} and formalize the solution that was presented informally. 
We first define the version corresponding to the example in \cref{sec:approach-overview} (\cref{sec:revisit-domains,sec:revisit-initiator,sec:revisit-filter-lenses}), and then extend it for the finer-grained updates presented in \cref{sec:finer-grained-descriptions-of-updates} (\cref{sec:elaborated-domains-for-finer-grained-updates}).

\newcommand{\FTables}{\con{Tasks}}
\newcommand{\MultiMap}[1]{#1^\mathrm{R}}
\newcommand{\DTables}{\con{DT}}
\newcommand{\DTablesOG}{\con{DT}_\mathrm{OG}}
\newcommand{\DTablesDT}{\con{DT}_\mathrm{DT}}

\newcommand{\DTablesS}{\con{DT}'}
\newcommand{\DTablesOGS}{\con{DT}_\mathrm{OG}'}
\newcommand{\DTablesDTS}{\con{DT}_\mathrm{DT}'}

\newcommand{\fto}{\rightharpoonup_\mathrm{f}}

Before defining concrete domains ($\FTables$, $\DTables$) involved in the transformation, we first present its overall structure as below. 
\begin{fullversionblock}
\[\bb
\ell_\mathrm{task} \in \MALens{\FTables}{(\DTables \x \DTables)}\\
\ell_\mathrm{task} = \TaskInit \fatsemi \DUP \fatsemi (\filterOnGoing \x \filterToday)
\ee\]   
\end{fullversionblock}

The above involves the following lenses and lens combinators (as well as $\fatsemi$).
{\myfootnotesize\begin{gather*}
\DUP \in \MALens{P}{(P \! \x \! P)} 
\quad{~~}
(\x) \in \MALens{P_1}{Q_1} \to \MALens{P_2}{Q_2} \to \MALens{(P_1 \!\x \!P_2)}{(Q_1 \!\x\! Q_2)}
\\
\TaskInit \in \MALens{\FTables}{\DTables} 
\quad{~~} \filterOnGoing \in \MALens{\DTables}{\DTables} 
\quad{~~} \filterToday   \in \MALens{\DTables}{\DTables}
\end{gather*}}%
Here, $P$ is an arbitrary duplicable \IPoset, and $P_1$, $P_2$, $Q_1$, and $Q_2$ are arbitrary \IPosets.
Among them, we have already seen the definition of the {\DUP}lication lens in \cref{sec:absence-aware-lenses} and confirmed its well-behavedness. In what follows,
after defining the domains involved, 
we will show concrete definitions of $\TaskInit$, $\filterOnGoing$, $\filterToday$, and $(\x)$, and confirm that the lenses ($\TaskInit$ and the filters) are well-behaved and
that the lens combinator $(\x)$ preserves well-behavedness.

\subsection{Domains}
\label{sec:revisit-domains}

\newcommand{\String}{\con{String}}
\newcommand{\Bool}{\con{Bool}}
\newcommand{\Date}{\con{Date}}
\newcommand{\FTablesOG}{\FTables_\mathrm{OG}}
\newcommand{\FTablesComp}{\FTables_\mathrm{Comp}}

Roughly speaking, 
$\FTables$ is a discrete poset of \fstates, and 
$\DTables$ additionally allows addition and deletion updates. 
For simplicity, we use these names to denote both the \IPosets and their underlying sets. 
Their definitions are similar to \cref{ex:partial-pstate}, but now we consider additions as well as deletions. 
We treat $t \in \FTables$ as a partial function, \ie, $\FTables \triangleq \con{ID} \pto (\Bool \x \String \x \Date)$. 
\Spstates have the form of $(A, D)$ with $A \in \FTables$ and $D \subseteq \con{ID}$, with the restriction of $\Disjoint(D,\dom(A))$. 
Here, we write $\Disjoint(S,T)$ to mean that $S$ and $T$ are disjoint, \ie, $S \cap T = \emptyset$.
Intuitively, $(A, D)$ constrains that $A$ must be present and $D$ must not be present;  $\Disjoint(D,\dom(A))$ ensures feasibility.
Formally, we define ${\DTables} \triangleq \FTables \cup \{ (A , D) \in \FTables \x 2^\con{ID} \mid \Disjoint(D, {\dom(A)}) \}$, with the 
following ordering and identical updates (obvious reflexivity rules are omitted). 
{\ifFullVersion\else\small\fi\begin{gather*}
\infer{ (A,D) \le (A', D') }{ A \subseteq A' \quad D \subseteq D'}
\quad 
\infer{ (A, D) \le t }{ A \subseteq t \quad \Disjoint(D,\dom(t)) }
\qquad 
\infer{ \IDon{(A,D)}{(A', D')} }{ A \subseteq A' \quad D \subseteq D'}
\quad 
\infer{ \IDon{(A,\emptyset)}{t} }{ A \subseteq t }
\end{gather*}}%
We define its merge operator $\oplus$ as below. 
{
\ifFullVersion
\begin{align*}
t \oplus t &= t \\ 
t \oplus (A, D) = (A, D) \oplus t &= t &&\text{if}~(A, D) \le t \\ 
(A,D) \oplus (A', D') &= (A \cup A', D \cup D')&&\text{if}~
 \bbt 
      (A \cup A') \in \FTables\\
      {} \wedge \Disjoint(D \cup D',\dom(A \cup A'))
 \ee
\end{align*}%
\else\small
\begin{gather*}
t \oplus t = t \qquad t \oplus (A, D) = (A, D) \oplus t = t~\text{if}~(A, D) \le t \\
(A,D) \oplus (A', D') = (A \cup A', D \cup D')~\text{if}~
 \bbt 
      (A \cup A') \in \FTables
      {} \wedge \Disjoint(D \cup D',\dom(A \cup A'))
 \ee
\end{gather*}%
\fi}%
The condition in the last line says that $(A \cup A', D \cup D')$ is a valid \pstate in \DTables.
This $\oplus$ soundly implements the join in \DTables, and is total and closed for $\IDonSet{x}$ for any $x$, which entails that $\DTables$ is duplicable. 
Actually, the construction of $\le$, $\IDonName$, and $\oplus$ here is an instance of the general recipe (\cref{sec:general-state-update-pairs}) and the conditions for duplicability are easy to enforce (\cref{lemma:generic-su-construction}).
This design of \IPoset is inspired by the 2P-Set CRDT~\cite{CRDT}, which uses a pair $(A,D)$ to represent the set $A \setminus D$.

An example of $\DTables$'s element is $w_\mathrm{merged}$ in \cref{sec:approach-overview}, where we used marks $(+)$ or $(-)$ in tables instead of using pairs $(A,D)$
The \pstates $\DStateOGVar$ and $\DStateDTVar$ are also elements in $\DTables$.

\newcommand{\SelOG}[1]{#1_\mathrm{OG}}
\newcommand{\SelDT}[1]{#1_\mathrm{DT}}

\subsection{\INITIATOR: \texorpdfstring{\TaskInit}{taskInit}}
\label{sec:revisit-initiator}

The \initiator $\TaskInit$ connects \fstates in $\FTables$ and \pstates in $\DTables$. 
We give it as an instance of a general form discussed in \cref{sec:initiators}. Since we have defined its source and view domains ($\FTables$ and $\DTables$), the remaining task is 
an appropriate definition of $(\ApplyUpd{}{})$ that satisfies \ref{law:u-acceptability} and \ref{law:u-consistency}. This is straightforward: 
\begin{fullversionblock}
\[\bbt
  \ApplyUpd{t'}{\dontcare} &= t' \qquad\qquad 
  \ApplyUpd{(A,D)}{t}      &= \left\{ (k \mapsto v) \in (t \lhd A) \mid k \not\in D \right\} %
\ee\]   
\end{fullversionblock}
{}
Here, $t \lhd A$ upserts $A$ into $t$\fullversion{, that is, it updates $t(k)$ with $A(k)$ or inserts $A(k)$ into $t$ for each $k \in \dom(A)$ depending on whether $t(k)$ is defined}. 
Formally, $(t \lhd A)(k)$ is defined as $A(k)$ if $k \in \dom(A)$ and otherwise as $t(k)$. 
For example, we have $\ApplyUpd{\DStateOGVar}{\OSourceVar} = \OSourceVar'$ and $\ApplyUpd{w_\mathrm{merged}}{\OSourceVar} = \OSourceVar''$. 

The order of upsertion and deletion above does not matter as we required $\Disjoint(D,\dom(A))$. Requiring this is a design choice: without this requirement, more \pstates can be merged, but the notion of preservation of updates is weakened. 
Without it, we need to give precedence for $A$ or $D$, which makes non-precedent one a soft constraint (as it may be overwritten).

\subsection{Filter Lenses}
\label{sec:revisit-filter-lenses}
Next, we give the definition of $\filterOnGoing \in \MALens{\DTables}{\DTables}$. We shall omit the one for \filterToday as it is similar. 
In the forward direction, the lens \filterOnGoing simply filters ongoing tasks. 
{\ifFullVersion\else\small\fi\[
\GET \; \filterOnGoing \; s = \begin{cases} 
  \SelOG{t} & \text{if}~s = t \in \FTables \\
  (\SelOG{A}, D) & \text{if}~s = (A, D) 
\end{cases}\]}%
Here, given $t \in \FTables$, we write $\SelOG{t}$ for a mapping obtained by collecting all ongoing tasks in $t$, \ie, $\SelOG{t} = \{ (k \mapsto v) \in t  \mid v = (\con{False}, \dontcare, \dontcare) \}$. %
In the backward direction, if the updated view is a \fstate, it must return a \fstate. Otherwise, the successive backward execution could add more tasks to violate the update preservation. Specifically, it upserts $t'$ into the original filtered out tasks. 
If the updated view is a \pstate, it just returns the \pstate intact regardless of the source $s$.
{\ifFullVersion\else\small\fi\[
\PUT \; \filterOnGoing \; (s, v) = \begin{cases}
   (t \setminus \SelOG{t}) \lhd t' & \text{if}~v = t' \in \FTablesOG \land s = t \in \FTables \\ 
   \left(A , D\right) &\text{if}~v = (A, D) \wedge A \in \FTablesOG\\
\end{cases} 
\]}%
We write $\FTablesOG$ for a set of \pstates that consists only of ongoing tasks, \ie, $\{ t \in \FTables \mid \forall k, t(k) = (\con{False}, \dontcare, \dontcare) \}$.
This $\PUT$ returns the updated view intact if it is a \spstate. An intuition behind this behavior is that a \spstate represents update requests and addition/removal requests on the filtered result can be translated into those on the original tasks.

This lens passes $D$ intact regardless of whether they are on ongoing tasks or not.
This is safe, in the sense that the lens is well-behaved\fullversion{ (see \cref{sec:well-behavedness-filterOnGoing-plain} for its proof)}.
For \ref{law:aa-consistency}, this only matters when we prove $(A, D) \le \GET \; \filterOnGoing \; t = \SelOG{t}$ from $\PUT \; \filterOnGoing \; (s, (A,D)) = (A,D) \le t$. 
Even when $D$ contains removal requests for completed tasks, we still have $(A,D) \le \SelOG{t}$ (provided that $A \in \FTablesOG$, which is ensured by the $\PUT$'s guard). Notice that, by $\SelOG{t} \subseteq t$, $\Disjoint(D,\dom(t))$ implies $\Disjoint(D, \dom(\SelOG{t}))$.
Recall that $D$ constrains that tasks with the IDs will not be present---it is allowed that more tasks will not be present (as long as $A$ is contained).
For \ref{law:aa-acceptability}, there is no such concern as $\IDon{(A, D)}{t}$ simply enforces $D$ to be the empty set.

\subsection{Lens Combinator \texorpdfstring{$\x$}{x}}

The definition of $\x$ is no different from the one appearing in the literature~\cite{PaCu10}.
Let $\ell_1 \in \MALens{P_1}{Q_1}$
and $\ell_2 \in \MALens{P_2}{Q_2}$ be lenses.
Then, we define the lens $(\ell_1 \times \ell_2) \in \MALens{(P_1 \times P_2)}{(Q_1 \times Q_2)}$, which applies $\ell_1$/$\ell_2$ to each component of the pair in parallel, as below.
{\ifFullVersion\else\small\fi\[
\bb
\GET \; (\ell_1 \times \ell_2) \; (s_1,s_2) = (\GET \; \ell_1 \; s_1, \GET \; \ell_2 \; s_2)\\
\PUT \; (\ell_1 \times \ell_2) \; ((s_1,s_2),(v_1,v_2)) =
 (\PUT \; \ell_1 \; (s_1,v_1), \PUT \; \ell_2 \; (s_2,v_2))
\ee
\]}%
Thanks to the point-wise ordering, the combinator $\times$ preserves well-behavedness.

\subsection{Elaborated Domains for Finer-Grained Updates} 
\label{sec:elaborated-domains-for-finer-grained-updates}

\cref{sec:finer-grained-descriptions-of-updates} mentioned that we can support finer-grained update descriptions (\spstates) $u_1, u_2$ that specifies the same possible results, \ie,
the set of \pstates $t$ with $u \le t$ is the same for $u = u_1, u_2$. 
The ability to treat such \pstates differently is one of the strengths of operation-based systems~\cite{Meertens98,HofmannPW12,DiskinXC11,AhmanU17}. Here we show that such control is also possible in our system. 
This is achieved by modifying the filter lenses in $\ell_\mathrm{task}$.

\subsubsection{Domains}

A natural approach is to make $\filterOnGoing$/$\filterToday$ convert completion/postponing requests to addition requests.
To make this possible, we use separate domains for filter results: $\DTablesOG$ and $\DTablesDT$, which support completion and postponing requests, respectively.  
Let us focus on $\DTablesOG$ as the definition of $\DTablesDT$ is similar. 
We define $\Delta_\mathrm{OG}$ as a set of triples $(A, C, D)$ where $A, C \in \FTables$
are sets of tasks, and $D \subseteq \con{ID}$ is a set of IDs, satisfying the following conditions:
\begin{fullversionblock}
\begin{itemize}
 \item $A(k) = (\con{False}, \dontcare, \dontcare)$ for all $k \in \dom(A)$,  
 \item $C(k) = (\con{True}, \dontcare, \dontcare)$ for all $k \in \dom(C)$, and 
 \item $\dom(A)$, $\dom(C)$ and $D$ are pairwise disjoint. 
\end{itemize}    
\end{fullversionblock} 

Intuitively, the first two conditions say that $A$ is an addition request for visible tasks in the view and $C$ is for invisible tasks. 
The third condition asserts the feasibility. 
Then, we define $\DTablesOG \triangleq \Delta_\mathrm{OG}$ with 
the following $\le$ and $\IDonName$.
{\ifFullVersion\else\small\fi\begin{gather*}
\infer{(A, C, D) \le (A' , C', D')}{A\subseteq A' \quad C \subseteq C' \quad D \subseteq D'}
\quad 
\infer{(A, C, D) \le t}{A \subseteq t \quad \Disjoint\left(\dom(C) \cup D,\dom(t)\right) }
\\[3pt plus 2pt]\displaybreak[0]
\infer{\IDon{(A,C,D)}{(A',C',D')}}{A\subseteq A' \quad C \subseteq C' \quad D \subseteq D'}
\quad 
\infer{\IDon{(A,\emptyset,\emptyset)}{t}}{A \subseteq t}   
\end{gather*}}%
In \cref{sec:finer-grained-descriptions-of-updates}, $A$, $C$, and $D$ are represented by $(+)$, $(\checkmark)$, and $(-)$ marks in the table. 
This $\DTablesOG$ is duplicable with $\oplus$ defined similarly to \cref{sec:revisit-domains}, but we shall omit the discussion as we do not use $\DUP$ for the domain.

\subsubsection{Modified $\filterOnGoing$}

The modification to $\filterOnGoing$ is straightforward. Now, its view domain is $\DTablesOG$ (and thus $\filterOnGoing \in \MALens{\DTables}{\DTablesOG}$).  
Its behavior on \spstates is changed: $\GET$ and $\PUT$ now interconvert
an addition request $A \uplus C$ in $\DTables$ with 
a pair of an addition request $A$ and a completion request $C$ in $\DTablesOG$, using the fact that $A \in \FTablesOG$ and $C \in \FTablesComp$.
{\ifFullVersion\else\small\fi\[
\bb
 \GET \; \filterOnGoing \; (A \uplus C, D) = (A, C, D) 
 ~\text{where}~A \in \FTablesOG, C \in \FTablesComp\\
 \PUT \; \filterOnGoing \; (\dontcare, (A,C,D)) = (A \uplus C, D)
 \ee
\]}%
(The definition of $\filterToday \in \MALens{\DTables}{\DTablesDT}$ is similar). 
The well-behavedness of this version is proved similarly to the previous version\fullversion{ (\cref{sec:well-behavedness-filterOnGoing})}. 
\section{\OUR{} Lenses as an Operation-Based System}
\label{sec:ours-as-operation-based-system}

This section shows that we can encode updates as \IPosets, providing a key advantage of operation-based systems: fine-grained control on backward behavior. 
\fullversion{The \IPosets that appear in this paper are based on the recipe with slight deviations.}

\subsection{State-Update Pairs}
\label{sec:general-state-update-pairs}

The basic idea to construct an \IPoset is to pair updates with their origins, 
inspired by encoding of operation-based CRDTs into state-based CRDTs~\cite{CRDT}.
Let $S$ be a set of (\fstateHead) states and $U$ be a poset of updates.\footnote{ 
Unlike Ahman and Uustalu~\cite{AhmanU17}, Diskin \etal~\cite{DiskinXC11}, we do not use a family of indexed sets to model updates, 
because we focus on their merging and applications. Using indexed sets makes the latter total, but merging is an intrinsically partial operation.
}
We assume that $U$ has a merge operator $\oplus_U$ that soundly implements the join in $U$, \ie, $u_1 \oplus_U u_2 = u$ implies $u_1 \vee u_2 = u$. We also assume that 
each $u \in U$ can be interpreted as a partial function $\SemUpd{u} \in S \pto S$, which intuitively represents the application of the update. 
Let us define $\Ran{s}{u} = \{ s' \in S \mid \exists u'.\, u \le_U u', \SemUpd{u'}(s) = s' \}$, which intuitively denotes a set of possible results of $u$ when the original state is $s$. 
Then, we define an \IPoset $\GenIPoset{S}{U} = (S \cup (S \times U), \le, \IDonName)$, where $\le$ and $\IDonName$ are given by: 
{\ifFullVersion\else\small\fi\[
\infer
{(s, u) \le (s, u')}
{ u \le_U u'}
\quad
\infer{s \le s}{} 
\quad 
\infer
{(s, u) \le s'}{s' \in \Ran{s}{u}}
\qquad 
\infer
{\IDon{(s, u)}{(s,u')}}
{ u \le_U u' }
\quad 
\infer
{\IDon{s}{s}}{}
\quad 
\infer 
{\IDon{(s,u)}{s}}
{\ApplyUpdD{u}{s} = s}
\]}%
Observe that $\le$ is a partial order and $\IDonName$ is a reflexive subset of $(\le)$. 

To support a \initiator, we then define $(\ApplyUpd{}{}) \in \GenIPoset{S}{U} \to S \pto S$ as\fullversion{ below. 
\[
\ApplyUpd{s}{\dontcare} = s \qquad \ApplyUpd{(s,u)}{s'} = \SemUpd{u}(s)~\text{if}~s = s' 
\]}{}
The \ref{law:u-consistency} and \ref{law:u-acceptability} of this $(\ApplyUpd{}{})$ follow %
from the definitions of $(\ApplyUpd{}{})$, $(\le)$, and $\IDonName$. 

Supporting $\DUP$ on $\GenIPoset{S}{U}$ requires some additional structures. 
We first define the merge operator $\oplus_{S,U} \in \GenIPoset{S}{U} \times \GenIPoset{S}{U} \pto \GenIPoset{S}{U}$ as below.
{\ifFullVersion\else\small\fi\begin{gather*}
s \oplus_{S,U} s' = s~\text{if}~s = s' 
\qquad s' \oplus_{S,U} (s,u) = (s,u) \oplus_{S,U} s' = s'~\text{if}~s' \in \Ran{s}{u} \\
(s,u) \oplus_{S,U} (s',u') = (s, u \oplus_U u')~\text{if}~s = s' 
\end{gather*}}%
Let us say that $\oplus_U$ respects $\Ran{s}{-}$ if $\Ran{s}{u_1} \cap \Ran{s}{u_2} \subseteq \Ran{s}{u}$ whenever $u_1 \oplus_U u_2 = u$. (The converse inclusion follows from the definition of $\Ran{s}{-}$.)

\newcommand{\Gconditions}{
 \item \label{cond:preserve-ran} $\oplus_U$ respects $\Ran{s}{-}$ for any $s$, 
 \item \label{cond:pijs} $\oplus_U$ is total on $\{ u' \mid u' \le_U u \}$ for any $u$, and 
 \item \label{cond:preserve-id} $\oplus_U$ is total and closed on $\{ u \mid \SemUpd{u}(s) = s \}$ for any $s$. 
}

\begin{lemma}
$\GenIPoset{S}{U}$ is duplicable with the merge operator $\oplus_{S,U}$ if the following conditions hold:%
\begin{fullversionblock}
\begin{enumerate}[label={\rm(G{\arabic*})}, leftmargin={32pt}]\Gconditions\end{enumerate}    
\end{fullversionblock}

\label{lemma:generic-su-construction}
\end{lemma}
\begin{fullversionblock}
\begin{proof}[Sketch] 
  We use \ref{cond:preserve-ran} to show that $\oplus_{S,U}$ soundly implements the join, and \ref{cond:pijs} and \ref{cond:preserve-id} to show that the merge operator is total and closed under $\IDonSet{x}$ (\ref{cond:pijs} for the $x = (s,u)$ case and \ref{cond:preserve-id} for the $x = s$ case). 
  More specifically, \ref{cond:preserve-ran} is used to ensure that $(s, u \oplus_U u')$ is the least element among the upper bounds of $(s, u)$ and $(s,u')$ in $\GenIPoset{S}{U}$. \qed 
\end{proof}  
\end{fullversionblock}

\begin{fullversionblock}

These conditions, \ref{cond:preserve-ran}, \ref{cond:pijs}, and \ref{cond:preserve-id}
are almost direct translations of the conditions required for $\GenIPoset{S}{U}$ to be duplicable. In fact, each of \ref{cond:preserve-ran}, \ref{cond:pijs}, and \ref{cond:preserve-id} is necessary.

We have the following sufficient condition for \ref{cond:preserve-ran} that does not refer to $\oplus_U$, which says that the update space $U$ must be fine enough in the sense that any two updates that have a common possible result can be refined into an update that also shares the possible result. 
\begin{lemma}
\ref{cond:preserve-ran} holds if, for all $s' \in \Ran{s}{u_1} \cap \Ran{s}{u_2}$, there exists $u$ such that $u_1 \le_U u$, $u_2 \le_U u$ and $s' \in \Ran{s}{u}$. \qed 
\label{lemma:ran-respected-for-fine-enough-updates}
\end{lemma}
\ref{cond:pijs} is implied by the following algebraic property of $\oplus_U$, which is desirable to have. 
\begin{lemma}
\ref{cond:pijs} holds if both of the following two conditions hold: 
(1) $\oplus_U$ is defined for any comparable $u_1$ and $u_2$, and 
(2) $\oplus_U$ is associative, \ie, $u_1 \oplus (u_2 \oplus u_3) = u \Leftrightarrow (u_1 \oplus u_2) \oplus u_3 = u$. \qed 
\label{lemma:pijs-holds-for-associatve-join}
\end{lemma}

\end{fullversionblock}

\subsection{Eliminating States from \Pstates}
Pairing a state $s$ with an update $u$ is inefficient as updates are typically smaller than states. 
Using $\Ran{}{u} = \bigcup_s \Ran{s}{u}$ in the premise of $u \le s'$ makes $\oplus_{S,U}$ unsound in general, as it violates \ref{cond:preserve-ran}, which is also a necessary condition for $\GenIPoset{S}{U}$ to be duplicable. 
However, this also means that 
$s$ can be omitted if either we will not use $\DUP$ for $\GenIPoset{S}{U}$, or $\oplus_{S,U}$ respects $\Ran{}{u} = \bigcup_s \Ran{s}{u}$. 
The \IPosets $\DTables$ and $\DTablesOG$ in \cref{sec:revisit-domains,sec:elaborated-domains-for-finer-grained-updates} are examples of the latter kind. 
They deviate slightly from the recipe; 
their $\IDonName$s are stricter than the recipe, and $\ApplyUpd{(A,D)}{t}$ is total, while 
the recipe assumes the partiality of $\SemUpd{(A,D)}$.

\begin{fullversionblock}
When $s \in \Ran{}{u}$ implies $\SemUpd{u}(s) = s$ for any $u$ in addition, we have $(\le) = \IDonName$. 
This explains why $\IDonName$ coincides with $(\le)$ in simple \IPosets, such as 
discrete ones (where $U$ is empty, and the condition is vacuously true), 
an \IPoset obtained from a discrete one by adding the distinct least element $\GO$ (where $\Ran{}{\GO} = S$ and $\SemUpd{\GO}(s) = s$ for any $s$), and the powerset \IPoset of $S$ with identification of \fstate $s$ with \pstate $\{s\}$ (where $\Ran{}{T} = T$ and $\SemUpd{T}(s) = s$ if $s \in T$, and otherwise $\SemUpd{T}(s) = t_s$ for a certain element $t_s \in T$).
\end{fullversionblock}

\section{Related Work}
\label{sec:related-work}

\Pstates have appeared in various forms in the bidirectional transformation literature.
The original lens formalization~\cite{FGMPS07,FosterGMPS05} uses $\GO$ to represent unavailable sources. 
This use of $\GO$, however, can be replaced by other means, such as the \var{create} function (\eg, \cite{BohannonFPPS08,HofmannPW11}) which computes the updated source only from the updated view.
A bidirectional programming language HOBiT~\cite{MaWa18esop} adopts \pstates for their treatment of value environments. 
Roughly speaking, they interpret a term-in-context $\GTH \vdash e : \sigma$ as a lens between value environments of type $\GTH$ and values of type ${\Gs}$, where its $\PUT$ execution specifies only the values of variables that occur in $e$.
\fullversion{For example, for $\GTH = \{ x : \con{Int}, y : \con{Bool} \}$, the $\PUT$ of $\GTH \vdash x : \con{Int}$ takes $1$ to return $\{ x \mapsto 1\}$,
and the $\PUT$ of $\GTH \vdash y : \con{Bool}$ takes $\con{True}$ to return $\{y \mapsto \con{True}\}$, specifying only the values of used variables.}
Such value environments can be seen as \pstates, which is crucial to 
interpret compound expressions like $(x,y)$ compositionally, where a compound expression can be seen as a microscopic form of multiple views. 
Their well-behavedness is compositional, partial-order based, and 
similar to our weak well-behavedness except for their maximality requirement, which ensures that their $\GET$ can only involve \fstates. 
They have no law corresponding to \ref{law:aa-stability} because their final source is essentially discrete due to external update reflection.
This difference would reflect where and how partial specifiedness is allowed: 
in their framework, \pstates are only permitted as the value environments and used to interpret terms-in-context.
We also note their goal is to guarantee classical well-behavedness~\cite{FosterGMPS05,FGMPS07}; they did not support view-to-view update propagation via source-sharing.

\newcommand{\ConsIns}{\mathbin{{{}^{+}{:}}}}
\newcommand{\ConsDel}{\mathbin{{{}^{-}{:}}}}

A similar idea appears in Mu \etal~\cite{MuHT04aplas}, where special tags are used to identify updated parts.
One form of tags is $*v$, which denotes that the part is updated and gives it precedence in the $\PUT$ execution of their duplication lens.
In this sense, their $*$-ed values are more specified than $*$-less values.
They also consider tags for structural list updates: $x \ConsIns \var{xs}$ and $x \ConsDel \var{xs}$ mark insertion and deletion of $x$, respectively.\footnote{Originally, the symbols $\oplus$ and $\ominus$ are used~\cite{MuHT04aplas}. We use different symbols to avoid confusion with our use of $\oplus$.}
The general recipe~(\cref{sec:general-state-update-pairs}) can handle their 
tagged lists with a slight modification. Since the origins of updates can be obtained from their tagged lists, the tagged lists can be regarded as state-and-update pairs. The ordering is then given by the backward behavior (say, $\var{merge}$) of their list duplication, namely, $x \le y \triangleq (\var{merge} \; x \; y = y)$. 
{
Note that we need to distinguish a list as a \fstate from that as a \pstate. The latter specifies more lists than itself; \eg, we have $[2,3] \le 1 \ConsIns [2,3]$. 
Since their $\var{merge}$ cannot merge identical updates such as $1 \ConsIns 1 \ConsDel [\,]$ and $1 \ConsDel 1 \ConsIns [\,]$, failing to satisfy \ref{cond:preserve-id}, we must further limit $\IDonName$ to accept only obviously identical updates like $1 : [\,]$, when they are compared with a \fstate. }
The removal of tags is performed by $\PUT$ of the \initiator.

\begin{fullversionblock}
The property called \textsc{WPutGet}~\cite{HidakaHIKMN10} is also used as a replacement of \ref{law:consistency} to characterize
the behavior of bidirectional transformations with the presence of multiple views%
\fullversion{.
\[
\PUT \; \ell \; (s_0, v) = s 
\; \Longrightarrow \; 
\PUT \; \ell \; (s_0, \GET \; \ell \; s) = s \tag{\textsc{WPutGet}}\label{law:wput-get}
\]}%

The idea of \textsc{WPutGet} is that, when there are multiple views,
a view update $v$ may be ``incomplete'' in the sense that it may not update all the copies consistently,
but we can adjust the updated view to $\GET \; \ell \; s$ so that all the copies will be updated consistently and that their backward propagation results will be identical.
The property is neither compositional~\cite[Appendix B]{MatsudaW15}, nor does it state the preservation of the user's updates.
We do not guarantee the property; actually, we have a well-behaved \our{} lens that does not satisfy \textsc{WPutGet} (See \cref{sec:appendix-lens-that-does-not-satisfy-WPutGet}).

As written in \cref{sec:preliminaries}, we do not require the {\ref{law:put-put}} law~\cite{FGMPS07} for our lenses%
{.
\[
 \PUT \; \ell \; (s_0, v_1) = s_1 \land \PUT \; \ell \; (s_1, v_2) = s_2 
 \Longrightarrow \PUT \; \ell \; (s_0, v_2) = s_2  \tag{\textsc{PutPut}}\label{law:put-put}
\]}
Classical lenses that satisfy this law in addition to \ref{law:consistency} and \ref{law:acceptability} are called very-well-behaved~\cite{FGMPS07}. 
Although {\ref{law:put-put}} is required in some existing frameworks~\cite{BancilhonS81,Hegner90,Hegner04,MatsudaHNHT07,Voigtlander09bff} for their construction of bidirectional transformations and is used as a basis for categorical generations of lenses~\cite{riley2018categoriesoptics,Boisseau20}, in the context of view updating, it is generally considered too strong to permit useful transformations~\cite{FGMPS07,Keller87}, such as projections for a database relation. 
In our framework, this law does not hold even for simple lenses such as \initiators, especially when the view allows ``no-op'' updates. 
For example, for a \initiator $\var{init} \in \MALens{\mathbb{N}}{\mathbb{N}_\GO}$, we have: $\PUT \; \var{init} \; (42, 1) = 1$ and $\PUT \; \var{init} \; (1, \GO) = 1$, while $\PUT \; \var{init} \; (42, \GO) = 42$.
\end{fullversionblock}

In the field of databases, Hegner~\cite{Hegner04} discusses laws for bidirectionality that refer to ordering.
Unlike ours, their ordering intuitively represents insertions and deletions.
Even for the non-order-related subpart, the laws are much stronger than the well-behavedness for classical lenses.
Besides \ref{law:acceptability} and \ref{law:consistency}, he also imposes
conditions that make $\PUT$ transparent~\cite{Hegner90}: intuitively, any updates can be undone, synchronization timing (invocation of $\PUT$) is irrelevant ({\ref{law:put-put}}), and whether $\PUT\; (s, v)$ is defined depends only on $\GET \; s$ and $v$.
By further requiring conditions on ordering, Hegner~\cite{Hegner04} shows the uniqueness of $\PUT$ for a given $\GET$.

\newcommand{\DPUT}{\mathit{dput}}
\newcommand{\DGET}{\mathit{dget}}
\newcommand{\EPUT}{\mathit{eput}}
\newcommand{\EGET}{\mathit{eget}}

\begin{fullversionblock}
The general recipe for encoding updates in \pstates (\cref{sec:general-state-update-pairs}) provides us a basis for direct comparisons with other operation-based frameworks. 
Delta lenses translate updates both forward and backward~\cite{DiskinXC11}.
Following the presentations in Ahman and Uustalu~\cite{AhmanU17}, 
a domain in delta lenses is a pair $(S, P)$ of a set $S$ of states and a family $\{P_s\}_{s \in S}$ of sets of updates that start from $s$. 
Then, a delta lens $\ell$ between $(S, P)$ and $(V, Q)$ is a triple of functions $\GET \; \ell \in S \to V$, $\DGET \; \ell \in (s : S) \to P \; s \to Q \; (\GET \; s)$, and $\DPUT \; \ell \in (s : S) \to Q \; (\GET \; s) \to P \; s$. Here, we used the Agda notation for dependent products. 
We would obtain similar laws even if we based our discussions on delta lenses. An advantage in this direction would be clear separation between \fstates and \spstates. For example, 
a form of well-behavedness focusing only on updates would be represented simply as a formula: 
$\DPUT \; \ell \; s \; u \le u' \Leftrightarrow u \le \DGET \; \ell \; s \; u'$. 
However, when $\DUP$ is concerned, this separation also makes it difficult to integrate the framework with classical lenses. 
Given that $(S, \lambda \dontcare.S)$ represents a state-based domain~\cite{AhmanU17}, 
a delta lens $\var{init}$ corresponding to \initiator will transform $(S, \lambda \dontcare.S)$ to $(S, P)$, for a certain $P$. It is natural that its $\DGET$ embeds a state $s'$ into a constant update ${!s'} \in P \; s$, \ie, $\DGET \; \var{init} \; s \; s' = {!s'}$, and its $\DPUT$ picks the update result, \ie, $\DPUT \; \var{init} \; s \; u = \SemUpd{u}(s)$. However, the unified law then enforces $\SemUpd{u}(s) = s' \Leftrightarrow u \le {!s'}$. 
This is too strong, as it requires that any comparable updates in $P \; s$ must have the same destination. For $u_1 \le u_2$, we have $u_2 \le {!}\SemUpd{u_2}(s)$ and hence $u_1 \le {!}\SemUpd{u_2}(s)$, which then implies $\SemUpd{u_1}(s) = \SemUpd{u_2}(s)$. 
This effectively limits the merge operator in $P \; s$ so that it can only merge updates with the same destinations.
Our framework avoids the issue by design: \fstates carry no origin information, and are handled differently in $\le$ and $\IDonName$ in the general recipe. 

Update-update lenses~\cite{AhmanU17} are a variant of delta lenses that can only translate updates backward.
Hence, it rules out domains that consist only of \spstates, such as CRDTs~\cite{CRDT,delta-CRDT}. 
Also, in the framework, we can only state the update preservation by comparing \spstates with \fstates. 
However, this fairly weakens the guarantee for lenses in the preservation of user's intentions; recall that updates carry more information than states in general. 
For example, consider $\DTablesOG$ in \cref{sec:elaborated-domains-for-finer-grained-updates}, where 
$\dom(C)$ and $D$ of $(A, C, D)$ are indistinguishable from a \fstate, \ie, 
$\{ t \mid (A, C, D) \le t \} = \{ t \mid (A, C', D') \le t \}$ when 
$\dom(C) \cup D = \dom(C') \cup D'$. 
It would be difficult for the update-update lens framework to reject lenses that garble such fine-grained intentions.

In edit lenses~\cite{HofmannPW12}, a lens retains an internal state $(\in C)$; its 
$\EGET \in (U_S, C) \to (U_V, C)$ and $\EPUT \in (U_V, C) \to (U_S, C)$ translate updates ($U_S$ on the source and $U_V$ on the view) between the source and view, while changing its internal state.\footnote{The original edit lens is an operation-based version of the symmetric lens~\cite{HofmannPW11}, but here we focus on the asymmetric case. Also, $\EGET$ and $\EPUT$ are not original names; they are called $\Rrightarrow$ and $\Lleftarrow$ originally~\cite{HofmannPW12}.}
A straightforward choice of $C$ is the source type $S$, when an edit lens can be expressed in terms of (a simply-typed version of) a delta lens as $\EGET \; (u,s) = (\DGET \; s \; u, \SemUpd{u}(s))$ and $\EPUT \; (u, s) = \LET~u' = \DPUT \; s \; u~\IN~(u', \SemUpd{u'}(s))$, but we often do not need the whole source information to translate updates; we also can store information that is lost during $\PUT$ in $C$.
Due to this flexibility of the internal state $C$, the discussion of laws in our setting would be more complex than for delta lenses.

These operation-based frameworks consider compositions of updates and require their update translation to preserve compositions. This is a form of the {\ref{law:put-put}}{} law~\cite{FGMPS07}, but unlike the state-based case, this requirement does not make their lenses impractical. In an extreme case where an update is given as a sequence of atomic updates, 
the composition preservation requirement is fulfilled simply by translating such atomic updates one-by-one. 
In our setting, however, %
the preservation of composition in their sense is less meaningful, 
because the updated view may differ from the view of the updated source, and hence, for a composition $u_1 \fatsemi u_2$ of updates, the view state after propagating $u_1$ may be different from the view that $u_2$ expects. 
We might express our operations in terms of compositions by treating $u_1 \le u_2$ as $u_2 = u_1 \fatsemi u$ for some $u$, and defining the merge operator of updates by its pushout. 
However, although this view would give a form of {\ref{law:put-put}}{} in our setting, this approach does not work well with a basic definition of updates, where updates are represented by sequences of primitive updates and different sequences are distinguished.
While one could consider an equational theory on such sequences to have meaningful pushouts, a more plausible solution is to assume a normal form of updates and perform operations on the normal forms, which is the approach adopted in this paper. 
\end{fullversionblock}

\section{Conclusion}
\label{sec:conclusion}

We introduced \emph{\our{} lenses}, extending the classical lenses~\cite{FosterGMPS05,FGMPS07} with \pstates. 
\Pstates are partially ordered, providing clear notions of update preservation and merging. This is crucial for handling multiple views that share a source. 
A key advantage of this framework is its support for compositional reasoning of order-aware well-behavedness, which is strong enough to guarantee the preservation of the user's updates. 

A future direction is to discuss more primitives and combinators for various datatypes, such as lists and finite maps. For lists, we would base our work on Mu \etal~\cite{MuHT04aplas}'s lists, as mentioned in \cref{sec:related-work}.
Another direction is to provide a frontend system so that users can express their intentions as \pstates. An approach would be to connect the proposed framework to an Elm-like architecture, providing a GUI to issue updates as \pstates.

\begin{credits}
\subsubsection{\ackname} We thank Zihang Ye, who inspired the direction of this research. 
This work was partially supported by JSPS KAKENHI Grant Numbers JP20H04161, JP23K20379, JP22H03562 and JP23K24818, and EPSRC Grant \emph{EXHIBIT: Expressive High-Level Languages for Bidirectional Transformations} (EP/T008911/1).

\subsubsection{\discintname}
The authors have no competing interests to declare that are
relevant to the content of this article. 
\end{credits}

\typeout{<<U: Current Page: \thepage>>}

\bibliographystyle{splncs04}

\begin{thebibliography}{10}
\providecommand{\url}[1]{\texttt{#1}}
\providecommand{\urlprefix}{URL }
\providecommand{\doi}[1]{https://doi.org/#1}

\bibitem{Abou-SalehCGMS16}
Abou{-}Saleh, F., Cheney, J., Gibbons, J., McKinna, J., Stevens, P.: Reflections on monadic lenses. In: Lindley, S., McBride, C., Trinder, P.W., Sannella, D. (eds.) A List of Successes That Can Change the World - Essays Dedicated to Philip Wadler on the Occasion of His 60th Birthday. Lecture Notes in Computer Science, vol.~9600, pp. 1--31. Springer (2016). \doi{10.1007/978-3-319-30936-1\_1}, \url{https://doi.org/10.1007/978-3-319-30936-1\_1}

\bibitem{AhmanU17}
Ahman, D., Uustalu, T.: Taking updates seriously. In: Eramo, R., Johnson, M. (eds.) Proceedings of the 6th International Workshop on Bidirectional Transformations co-located with The European Joint Conferences on Theory and Practice of Software, BX@ETAPS 2017, Uppsala, Sweden, April 29, 2017. {CEUR} Workshop Proceedings, vol.~1827, pp. 59--73. CEUR-WS.org (2017), \url{https://ceur-ws.org/Vol-1827/paper11.pdf}

\bibitem{delta-CRDT}
Almeida, P.S., Shoker, A., Baquero, C.: Delta state replicated data types. J. Parallel Distributed Comput.  \textbf{111},  162--173 (2018). \doi{10.1016/J.JPDC.2017.08.003}, \url{https://doi.org/10.1016/j.jpdc.2017.08.003}

\bibitem{BancilhonS81}
Bancilhon, F., Spyratos, N.: Update semantics of relational views. {ACM} Trans. Database Syst.  \textbf{6}(4),  557--575 (1981). \doi{10.1145/319628.319634}, \url{https://doi.org/10.1145/319628.319634}

\bibitem{BarbosaCFGP10}
Barbosa, D.M.J., Cretin, J., Foster, N., Greenberg, M., Pierce, B.C.: Matching lenses: alignment and view update. In: Hudak, P., Weirich, S. (eds.) Proceeding of the 15th {ACM} {SIGPLAN} international conference on Functional programming, {ICFP} 2010, Baltimore, Maryland, USA, September 27-29, 2010. pp. 193--204. {ACM} (2010). \doi{10.1145/1863543.1863572}, \url{https://doi.org/10.1145/1863543.1863572}

\bibitem{BohannonFPPS08}
Bohannon, A., Foster, J.N., Pierce, B.C., Pilkiewicz, A., Schmitt, A.: Boomerang: resourceful lenses for string data. In: Necula, G.C., Wadler, P. (eds.) Proceedings of the 35th {ACM} {SIGPLAN-SIGACT} Symposium on Principles of Programming Languages, {POPL} 2008, San Francisco, California, USA, January 7-12, 2008. pp. 407--419. {ACM} (2008). \doi{10.1145/1328438.1328487}, \url{https://doi.org/10.1145/1328438.1328487}

\bibitem{BohannonPV06}
Bohannon, A., Pierce, B.C., Vaughan, J.A.: Relational lenses: a language for updatable views. In: Vansummeren, S. (ed.) Proceedings of the Twenty-Fifth {ACM} {SIGACT-SIGMOD-SIGART} Symposium on Principles of Database Systems, June 26-28, 2006, Chicago, Illinois, {USA}. pp. 338--347. {ACM} (2006). \doi{10.1145/1142351.1142399}, \url{https://doi.org/10.1145/1142351.1142399}

\bibitem{Boisseau20}
Boisseau, G.: String diagrams for optics. In: Ariola, Z.M. (ed.) 5th International Conference on Formal Structures for Computation and Deduction, {FSCD} 2020, June 29-July 6, 2020, Paris, France (Virtual Conference). LIPIcs, vol.~167, pp. 17:1--17:18. Schloss Dagstuhl - Leibniz-Zentrum f{\"{u}}r Informatik (2020). \doi{10.4230/LIPICS.FSCD.2020.17}, \url{https://doi.org/10.4230/LIPIcs.FSCD.2020.17}

\bibitem{Chin93}
Chin, W.: Towards an automated tupling strategy. In: Schmidt, D.A. (ed.) Proceedings of the {ACM} {SIGPLAN} Symposium on Partial Evaluation and Semantics-Based Program Manipulation, PEPM'93, Copenhagen, Denmark, June 14-16, 1993. pp. 119--132. {ACM} (1993). \doi{10.1145/154630.154643}, \url{https://doi.org/10.1145/154630.154643}

\bibitem{CunhaFMPS12}
Cunha, J., Fernandes, J.P., Mendes, J., Pacheco, H., Saraiva, J.: Bidirectional transformation of model-driven spreadsheets. In: Hu, Z., de~Lara, J. (eds.) Theory and Practice of Model Transformations - 5th International Conference, ICMT@TOOLS 2012, Prague, Czech Republic, May 28-29, 2012. Proceedings. Lecture Notes in Computer Science, vol.~7307, pp. 105--120. Springer (2012). \doi{10.1007/978-3-642-30476-7\_7}, \url{https://doi.org/10.1007/978-3-642-30476-7\_7}

\bibitem{DiskinXC11}
Diskin, Z., Xiong, Y., Czarnecki, K.: From state- to delta-based bidirectional model transformations: the asymmetric case. J. Object Technol.  \textbf{10},  6: 1--25 (2011). \doi{10.5381/JOT.2011.10.1.A6}, \url{https://doi.org/10.5381/jot.2011.10.1.a6}

\bibitem{FischerHP15}
Fischer, S., Hu, Z., Pacheco, H.: A clear picture of lens laws - functional pearl. In: Hinze, R., Voigtl{\"{a}}nder, J. (eds.) Mathematics of Program Construction - 12th International Conference, {MPC} 2015, K{\"{o}}nigswinter, Germany, June 29 - July 1, 2015. Proceedings. Lecture Notes in Computer Science, vol.~9129, pp. 215--223. Springer (2015). \doi{10.1007/978-3-319-19797-5\_10}, \url{https://doi.org/10.1007/978-3-319-19797-5\_10}

\bibitem{FosterGMPS05}
Foster, J.N., Greenwald, M.B., Moore, J.T., Pierce, B.C., Schmitt, A.: Combinators for bi-directional tree transformations: a linguistic approach to the view update problem. In: Palsberg, J., Abadi, M. (eds.) Proceedings of the 32nd {ACM} {SIGPLAN-SIGACT} Symposium on Principles of Programming Languages, {POPL} 2005, Long Beach, California, USA, January 12-14, 2005. pp. 233--246. {ACM} (2005). \doi{10.1145/1040305.1040325}, \url{https://doi.org/10.1145/1040305.1040325}

\bibitem{FGMPS07}
Foster, J.N., Greenwald, M.B., Moore, J.T., Pierce, B.C., Schmitt, A.: Combinators for bidirectional tree transformations: A linguistic approach to the view-update problem. ACM Trans. Program. Lang. Syst.  \textbf{29}(3) (2007)

\bibitem{FosterPP08}
Foster, J.N., Pilkiewicz, A., Pierce, B.C.: Quotient lenses. In: Hook, J., Thiemann, P. (eds.) Proceeding of the 13th {ACM} {SIGPLAN} international conference on Functional programming, {ICFP} 2008, Victoria, BC, Canada, September 20-28, 2008. pp. 383--396. {ACM} (2008). \doi{10.1145/1411204.1411257}, \url{https://doi.org/10.1145/1411204.1411257}

\bibitem{GoldsteinFWP23}
Goldstein, H., Frohlich, S., Wang, M., Pierce, B.C.: Reflecting on random generation. Proc. {ACM} Program. Lang.  \textbf{7}({ICFP}),  322--355 (2023). \doi{10.1145/3607842}, \url{https://doi.org/10.1145/3607842}

\bibitem{Hegner90}
Hegner, S.J.: Foundations of canonical update support for closed database views. In: Abiteboul, S., Kanellakis, P.C. (eds.) ICDT'90, Third International Conference on Database Theory, Paris, France, December 12-14, 1990, Proceedings. Lecture Notes in Computer Science, vol.~470, pp. 422--436. Springer (1990). \doi{10.1007/3-540-53507-1\_93}, \url{https://doi.org/10.1007/3-540-53507-1\_93}

\bibitem{Hegner04}
Hegner, S.J.: An order-based theory of updates for closed database views. Ann. Math. Artif. Intell.  \textbf{40}(1-2),  63--125 (2004). \doi{10.1023/A:1026158013113}, \url{https://doi.org/10.1023/A:1026158013113}

\bibitem{HidakaHIKMN10}
Hidaka, S., Hu, Z., Inaba, K., Kato, H., Matsuda, K., Nakano, K.: Bidirectionalizing graph transformations. In: Hudak, P., Weirich, S. (eds.) Proceeding of the 15th {ACM} {SIGPLAN} international conference on Functional programming, {ICFP} 2010, Baltimore, Maryland, USA, September 27-29, 2010. pp. 205--216. {ACM} (2010). \doi{10.1145/1863543.1863573}, \url{https://doi.org/10.1145/1863543.1863573}

\bibitem{HofmannPW11}
Hofmann, M., Pierce, B.C., Wagner, D.: Symmetric lenses. In: Ball, T., Sagiv, M. (eds.) Proceedings of the 38th {ACM} {SIGPLAN-SIGACT} Symposium on Principles of Programming Languages, {POPL} 2011, Austin, TX, USA, January 26-28, 2011. pp. 371--384. {ACM} (2011). \doi{10.1145/1926385.1926428}, \url{https://doi.org/10.1145/1926385.1926428}

\bibitem{HofmannPW12}
Hofmann, M., Pierce, B.C., Wagner, D.: Edit lenses. In: Field, J., Hicks, M. (eds.) Proceedings of the 39th {ACM} {SIGPLAN-SIGACT} Symposium on Principles of Programming Languages, {POPL} 2012, Philadelphia, Pennsylvania, USA, January 22-28, 2012. pp. 495--508. {ACM} (2012). \doi{10.1145/2103656.2103715}, \url{https://doi.org/10.1145/2103656.2103715}

\bibitem{HornPC18}
Horn, R., Perera, R., Cheney, J.: Incremental relational lenses. Proc. {ACM} Program. Lang.  \textbf{2}({ICFP}),  74:1--74:30 (2018). \doi{10.1145/3236769}, \url{https://doi.org/10.1145/3236769}

\bibitem{HuITT97}
Hu, Z., Iwasaki, H., Takeichi, M., Takano, A.: Tupling calculation eliminates multiple data traversals. In: {Peyton Jones}, S.L., Tofte, M., Berman, A.M. (eds.) Proceedings of the 1997 {ACM} {SIGPLAN} International Conference on Functional Programming {(ICFP} '97), Amsterdam, The Netherlands, June 9-11, 1997. pp. 164--175. {ACM} (1997). \doi{10.1145/258948.258964}, \url{https://doi.org/10.1145/258948.258964}

\bibitem{HuK16}
Hu, Z., Ko, H.: Principles and practice of bidirectional programming in bigul. In: Gibbons, J., Stevens, P. (eds.) Bidirectional Transformations - International Summer School, Oxford, UK, July 25-29, 2016, Tutorial Lectures. Lecture Notes in Computer Science, vol.~9715, pp. 100--150. Springer (2016). \doi{10.1007/978-3-319-79108-1\_4}, \url{https://doi.org/10.1007/978-3-319-79108-1\_4}

\bibitem{HuMT04}
Hu, Z., Mu, S., Takeichi, M.: A programmable editor for developing structured documents based on bidirectional transformations. In: Heintze, N., Sestoft, P. (eds.) Proceedings of the 2004 {ACM} {SIGPLAN} Workshop on Partial Evaluation and Semantics-based Program Manipulation, 2004, Verona, Italy, August 24-25, 2004. pp. 178--189. {ACM} (2004). \doi{10.1145/1014007.1014025}, \url{https://doi.org/10.1145/1014007.1014025}

\bibitem{Keller87}
Keller, A.M.: Comments on bancilhon and spyratos' "update semantics and relational views". {ACM} Trans. Database Syst.  \textbf{12}(3),  521--523 (1987). \doi{10.1145/27629.214296}, \url{https://doi.org/10.1145/27629.214296}

\bibitem{KoZH16}
Ko, H., Zan, T., Hu, Z.: Bigul: a formally verified core language for putback-based bidirectional programming. In: Erwig, M., Rompf, T. (eds.) Proceedings of the 2016 {ACM} {SIGPLAN} Workshop on Partial Evaluation and Program Manipulation, {PEPM} 2016, St. Petersburg, FL, USA, January 20 - 22, 2016. pp. 61--72. {ACM} (2016). \doi{10.1145/2847538.2847544}, \url{http://doi.acm.org/10.1145/2847538.2847544}

\bibitem{vanLaarhovenLens}
van Laarhoven, T.: Cps based functional references. blog post: \url{https://www.twanvl.nl/blog/haskell/cps-functional-references} (2009), visited 2024-10-08

\bibitem{MacedoPSC14}
Macedo, N., Pacheco, H., Sousa, N.R., Cunha, A.: Bidirectional spreadsheet formulas. In: Fleming, S.D., Fish, A., Scaffidi, C. (eds.) {IEEE} Symposium on Visual Languages and Human-Centric Computing, {VL/HCC} 2014, Melbourne, VIC, Australia, July 28 - August 1, 2014. pp. 161--168. {IEEE} Computer Society (2014). \doi{10.1109/VLHCC.2014.6883041}, \url{https://doi.org/10.1109/VLHCC.2014.6883041}

\bibitem{MatsudaHNHT07}
Matsuda, K., Hu, Z., Nakano, K., Hamana, M., Takeichi, M.: Bidirectionalization transformation based on automatic derivation of view complement functions. In: Hinze, R., Ramsey, N. (eds.) Proceedings of the 12th {ACM} {SIGPLAN} International Conference on Functional Programming, {ICFP} 2007, Freiburg, Germany, October 1-3, 2007. pp. 47--58. {ACM} (2007). \doi{10.1145/1291151.1291162}, \url{https://doi.org/10.1145/1291151.1291162}

\bibitem{MaWa13}
Matsuda, K., Wang, M.: {FliPpr}: {A} prettier invertible printing system. In: Felleisen, M., Gardner, P. (eds.) ESOP. Lecture Notes in Computer Science, vol.~7792, pp. 101--120. Springer (2013). \doi{10.1007/978-3-642-37036-6_6}, \url{https://doi.org/10.1007/978-3-642-37036-6_6}

\bibitem{MaWa15}
Matsuda, K., Wang, M.: Applicative bidirectional programming with lenses. In: Fisher, K., Reppy, J.H. (eds.) {ICFP}. pp. 62--74. {ACM} (2015). \doi{10.1145/2784731.2784750}, \url{http://doi.acm.org/10.1145/2784731.2784750}

\bibitem{MatsudaW15}
Matsuda, K., Wang, M.: ``{B}idirectionalization for free'' for monomorphic transformations. Sci. Comput. Program.  \textbf{111},  79--109 (2015). \doi{10.1016/j.scico.2014.07.008}, \url{https://doi.org/10.1016/j.scico.2014.07.008}

\bibitem{MaWa18jfp}
Matsuda, K., Wang, M.: Applicative bidirectional programming: Mixing lenses and semantic bidirectionalization. J. Funct. Program.  \textbf{28}, ~e15 (2018). \doi{10.1017/S0956796818000096}, \url{https://doi.org/10.1017/S0956796818000096}

\bibitem{MatsudaW18haskell}
Matsuda, K., Wang, M.: Embedding invertible languages with binders: a case of the {FliPpr} language. In: Wu, N. (ed.) Proceedings of the 11th {ACM} {SIGPLAN} International Symposium on Haskell, Haskell@ICFP 2018, St. Louis, MO, USA, September 27-17, 2018. pp. 158--171. {ACM} (2018). \doi{10.1145/3242744.3242758}, \url{https://doi.org/10.1145/3242744.3242758}

\bibitem{MaWa18esop}
Matsuda, K., Wang, M.: {HOBiT}: Programming lenses without using lens combinators. In: Ahmed, A. (ed.) {ESOP}. Lecture Notes in Computer Science, vol. 10801, pp. 31--59. Springer (2018). \doi{10.1007/978-3-319-89884-1_2}, \url{https://doi.org/10.1007/978-3-319-89884-1_2}

\bibitem{MayerKC18}
Mayer, M., Kuncak, V., Chugh, R.: Bidirectional evaluation with direct manipulation. Proc. {ACM} Program. Lang.  \textbf{2}({OOPSLA}),  127:1--127:28 (2018). \doi{10.1145/3276497}, \url{https://doi.org/10.1145/3276497}

\bibitem{Meertens98}
Meertens, L.: Designing constraint maintainers for user interaction (1998), available on: \url{https://www.kestrel.edu/people/meertens/pub/dcm.pdf}

\bibitem{MuHT04aplas}
Mu, S., Hu, Z., Takeichi, M.: An algebraic approach to bi-directional updating. In: Chin, W. (ed.) Programming Languages and Systems: Second Asian Symposium, {APLAS} 2004, Taipei, Taiwan, November 4-6, 2004. Proceedings. Lecture Notes in Computer Science, vol.~3302, pp. 2--20. Springer (2004). \doi{10.1007/978-3-540-30477-7\_2}, \url{https://doi.org/10.1007/978-3-540-30477-7\_2}

\bibitem{OConnor11}
O'Connor, R.: Functor is to lens as applicative is to biplate: Introducing multiplate. CoRR  \textbf{abs/1103.2841} (2011), \url{http://arxiv.org/abs/1103.2841}, accepted in WGP '11, but not included in its proceedings

\bibitem{PaCu10}
Pacheco, H., Cunha, A.: Generic point-free lenses. In: Bolduc, C., Desharnais, J., Ktari, B. (eds.) Mathematics of Program Construction, 10th International Conference, {MPC} 2010, Qu{\'{e}}bec City, Canada, June 21-23, 2010. Proceedings. Lecture Notes in Computer Science, vol.~6120, pp. 331--352. Springer (2010). \doi{10.1007/978-3-642-13321-3\_19}, \url{https://doi.org/10.1007/978-3-642-13321-3\_19}

\bibitem{PickeringGW17}
Pickering, M., Gibbons, J., Wu, N.: Profunctor optics: Modular data accessors. Art Sci. Eng. Program.  \textbf{1}(2), ~7 (2017). \doi{10.22152/PROGRAMMING-JOURNAL.ORG/2017/1/7}, \url{https://doi.org/10.22152/programming-journal.org/2017/1/7}

\bibitem{riley2018categoriesoptics}
Riley, M.: Categories of optics (2018), \url{https://arxiv.org/abs/1809.00738}

\bibitem{CRDT}
Shapiro, M., Pregui{\c{c}}a, N.M., Baquero, C., Zawirski, M.: Conflict-free replicated data types. In: D{\'{e}}fago, X., Petit, F., Villain, V. (eds.) Stabilization, Safety, and Security of Distributed Systems - 13th International Symposium, {SSS} 2011, Grenoble, France, October 10-12, 2011. Proceedings. Lecture Notes in Computer Science, vol.~6976, pp. 386--400. Springer (2011). \doi{10.1007/978-3-642-24550-3\_29}, \url{https://doi.org/10.1007/978-3-642-24550-3\_29}

\bibitem{Stevens07}
Stevens, P.: A landscape of bidirectional model transformations. In: L{\"{a}}mmel, R., Visser, J., Saraiva, J. (eds.) GTTSE. Lecture Notes in Computer Science, vol.~5235, pp. 408--424. Springer (2008). \doi{10.1007/978-3-540-88643-3_10}, \url{https://doi.org/10.1007/978-3-540-88643-3_10}

\bibitem{Takeichi2009}
Takeichi, M.: Configuring bidirectional programs with functions. Presented at IFL 2009: International Symposium/Workshop on Implementation and Application of Functional Languages (2009), available from: \url{https://takeichimasato.net/attachments/XFun0.pdf}

\bibitem{Voigtlander09bff}
Voigtl{\"{a}}nder, J.: Bidirectionalization for free! (pearl). In: Shao, Z., Pierce, B.C. (eds.) Proceedings of the 36th {ACM} {SIGPLAN-SIGACT} Symposium on Principles of Programming Languages, {POPL} 2009, Savannah, GA, USA, January 21-23, 2009. pp. 165--176. {ACM} (2009). \doi{10.1145/1480881.1480904}, \url{https://doi.org/10.1145/1480881.1480904}

\bibitem{VoigtlanderHMW10}
Voigtl{\"{a}}nder, J., Hu, Z., Matsuda, K., Wang, M.: Combining syntactic and semantic bidirectionalization. In: Hudak, P., Weirich, S. (eds.) Proceeding of the 15th {ACM} {SIGPLAN} international conference on Functional programming, {ICFP} 2010, Baltimore, Maryland, USA, September 27-29, 2010. pp. 181--192. {ACM} (2010). \doi{10.1145/1863543.1863571}, \url{https://doi.org/10.1145/1863543.1863571}

\bibitem{VoigtlanderHMW13}
Voigtl{\"{a}}nder, J., Hu, Z., Matsuda, K., Wang, M.: Enhancing semantic bidirectionalization via shape bidirectionalizer plug-ins. J. Funct. Program.  \textbf{23}(5),  515--551 (2013). \doi{10.1017/S0956796813000130}, \url{https://doi.org/10.1017/S0956796813000130}

\bibitem{WilliamsG22}
Williams, J., Gordon, A.D.: Where-provenance for bidirectional editing in spreadsheets. J. Comput. Lang.  \textbf{73},  101155 (2022). \doi{10.1016/J.COLA.2022.101155}, \url{https://doi.org/10.1016/j.cola.2022.101155}

\bibitem{XieSH25}
Xie, R., Schrijvers, T., Hu, Z.: Biparsers: Exact printing for data synchronisation. Proc. {ACM} Program. Lang.  \textbf{9}({POPL}),  2205--2231 (2025). \doi{10.1145/3704910}, \url{https://doi.org/10.1145/3704910}

\bibitem{XiongLHZTM07}
Xiong, Y., Liu, D., Hu, Z., Zhao, H., Takeichi, M., Mei, H.: Towards automatic model synchronization from model transformations. In: Stirewalt, R.E.K., Egyed, A., Fischer, B. (eds.) 22nd {IEEE/ACM} International Conference on Automated Software Engineering {(ASE} 2007), November 5-9, 2007, Atlanta, Georgia, {USA}. pp. 164--173. {ACM} (2007). \doi{10.1145/1321631.1321657}, \url{https://doi.org/10.1145/1321631.1321657}

\bibitem{YuLHHKM12}
Yu, Y., Lin, Y., Hu, Z., Hidaka, S., Kato, H., Montrieux, L.: Maintaining invariant traceability through bidirectional transformations. In: Glinz, M., Murphy, G.C., Pezz{\`{e}}, M. (eds.) 34th International Conference on Software Engineering, {ICSE} 2012, June 2-9, 2012, Zurich, Switzerland. pp. 540--550. {IEEE} Computer Society (2012). \doi{10.1109/ICSE.2012.6227162}, \url{https://doi.org/10.1109/ICSE.2012.6227162}

\bibitem{ZhangH22}
Zhang, X., Hu, Z.: Towards bidirectional live programming for incomplete programs. In: 44th {IEEE/ACM} 44th International Conference on Software Engineering, {ICSE} 2022, Pittsburgh, PA, USA, May 25-27, 2022. pp. 2154--2164. {ACM} (2022). \doi{10.1145/3510003.3510195}, \url{https://doi.org/10.1145/3510003.3510195}

\bibitem{ZhangXGHZH24}
Zhang, X., Xie, R., Guo, G., He, X., Zan, T., Hu, Z.: Fusing direct manipulations into functional programs. Proc. {ACM} Program. Lang.  \textbf{8}({POPL}),  1211--1238 (2024). \doi{10.1145/3632883}, \url{https://doi.org/10.1145/3632883}

\bibitem{ZhuK0SH15}
Zhu, Z., Ko, H., Martins, P., Saraiva, J., Hu, Z.: Biyacc: Roll your parser and reflective printer into one. In: Cunha, A., Kindler, E. (eds.) Proceedings of the 4th International Workshop on Bidirectional Transformations co-located with Software Technologies: Applications and Foundations, {STAF} 2015, L'Aquila, Italy, July 24, 2015. {CEUR} Workshop Proceedings, vol.~1396, pp. 43--50. CEUR-WS.org (2015), \url{https://ceur-ws.org/Vol-1396/p43-zhu.pdf}

\bibitem{ZhuZK0SH16}
Zhu, Z., Zhang, Y., Ko, H., Martins, P., Saraiva, J., Hu, Z.: Parsing and reflective printing, bidirectionally. In: van~der Storm, T., Balland, E., Varr{\'{o}}, D. (eds.) Proceedings of the 2016 {ACM} {SIGPLAN} International Conference on Software Language Engineering, Amsterdam, The Netherlands, October 31 - November 1, 2016. pp. 2--14. {ACM} (2016). \doi{10.1145/2997364}, \url{http://dl.acm.org/citation.cfm?id=2997369}

\end{thebibliography}

\ifFullVersion
\clearpage 
\typeout{<<U: Appendix Starts: \thepage>>}

\appendix
\renewcommand\theHsection{\thesection.appendix}
\crefalias{section}{appendix}
\crefalias{subsection}{appendix}

\section{Proofs}
\label{sec:proofs}

\subsection{\cref{lemma:get-monotone}}
\label{sec:proof-get-monotone}

Let $v$ be $\GET \; \ell \; s$. Suppose $s \le s'$. By \ref{law:aa-acceptability}, $\PUT \; \ell \; (s,v)$ is defined and yields some $s''$ such that $\IDon{s''}{s}$. Since $\IDonName$ is a subset of $\le$, we then have $s'' \le s$. 
Then, by \ref{law:aa-consistency} with $s'' \le s \le s'$, we have $v \le \GET \; \ell \; s'$.

\subsection{\cref{lemma:view-stability}}
\label{sec:proof-view-stability}

Let $v$ be $\GET \; \ell \; s$.

By \ref{law:aa-acceptability}, $\PUT \; \ell \; (s, v)$ must be defined; let the result be $s'$. It also ensures $\IDon{s'}{s}$, which implies $s' \le s$. 
Then, we apply \ref{law:aa-consistency} to $\PUT \; \ell \; (s, v) = s'$ and $s' \le s'$, 
which yields $v \le \GET \; \ell \; s'$. 
Next, by the monotonicity of $\GET$ (\cref{lemma:get-monotone}) with $s' \le s$, 
we have $\GET \; \ell \; s' \le \GET \; \ell \; s = v$. 
Since we have both $v \le \GET \; \ell \; s'$ and $\GET \; \ell \; s' \le v$, 
we conclude $\GET \; \ell \; s' = v$ by antisymmetry.

\subsection{\cref{lemma:put-determines-get}}
\label{sec:proof-put-determines-get}

We first prove $v \le \GET \; \ell \; s$ for any $v \in V_s$. %
and then prove $\GET \; \ell \; s \in V_s$. %
Let $v$ be an element of $V_s$, meaning that we have $\PUT \; \ell \; (s_0, v) = s_1$ and $s_1 \le s$ for some $s_0$ and $s_1$. Then, by \ref{law:aa-consistency}, we have $v \le \GET \; \ell \; s$. %
Next, let $\GET \; \ell \; s$ be $v$. By \ref{law:aa-acceptability}, we have $\PUT \; \ell \;  (s, v) = s'$ for some $s'$ with $\IDon{s'}{s}$, implying $s' \le s$. Hence, we have $\GET \; \ell \; s \in V_s$.

\subsection{Well-Behavedness of \texorpdfstring{$\filterOnGoing$}{filterOnGoing} in \cref{sec:revisit-filter-lenses}}
\label{sec:well-behavedness-filterOnGoing-plain}

We note that $\filterOnGoing$ is defined so that it is classically well-behaved when only \fstates are involved, and thus 
we only need to examine the case where \spstates are involved. 

First, we confirm \ref{law:aa-consistency}.
Suppose that $\PUT \; \filterOnGoing \; (s,v) = s'$. Since $\GET \; \filterOnGoing$ is monotone, it suffices to show $v \le \GET \; \filterOnGoing \; s'$. 
By case analysis on $v$, it is easy to see $v = \GET \; \filterOnGoing \; s'$ by definition of $\filterOnGoing$. 

Second, we confirm \ref{law:aa-acceptability}. 
Suppose that $\GET \; \filterOnGoing \; s = v$ and $\IDon{v'}{v}$. 
We will show $\PUT \; \filterOnGoing \; (s,v') = s'$ for some $s'$ with $\IDon{s'}{s}$. 
Consider the case where $s = t \in \FTables$. If $v' = v = \SelOG{t}$, then it is easy to see that $s' = (t \setminus \SelOG{t}) \lhd \SelOG{t} = t$, ensuring $s' = \IDon{t}{t}$. 
If $v' = (A, D) \in \IDonSet{\SelOG{t}}$, then it must be the case where $A \subseteq \SelOG{t} \in \FTablesOG$ and $D = \emptyset$, which implies $s' = (A,\emptyset)$. 
Since we have $A \subseteq \SelOG{t} \subseteq t$, we have $s' = (A, \emptyset) \in \IDonSet{t}$.
Consider the case where $s = (A,D)$. Then, it must be the case where $v = (\SelOG{A}, D)$. 
By definition, $v'$ that yields the greatest $s'$ is $v' = v$, and it suffices to consider the case where $v' = v$. 
In this case, $s' = (\SelOG{A}, D)$. Since we have $\SelOG{A} \subseteq A$, we have $s' = \IDon{(\SelOG{A},D)}{(A,D)}$.

Finally, we confirm \ref{law:aa-stability}. 
Suppose that $\PUT \; \filterOnGoing \; (s_0,v) = s$, and $\PUT \; \filterOnGoing \; (s', v'') = s''$ with $s \le s'$ and $v \le v'' \in \IDonSet{\GET \; \filterOnGoing \; s'}$. We will show $s \le s''$. When $s$ is a \fstate, only \fstates are involved in the discussion. 
Hence, we consider the case where $s = v = (A,D)$ with $A \in \FTablesOG$. 
When $s' = (A', D')$, $\GET \; \filterOnGoing \; s'$ is also a \spstate. 
For \spstates $v_1, v_2$ with $v_1 \le v_2$, we have $\PUT \; \filterOnGoing \; (s', v_2) = s_2$ implies $\PUT \; \filterOnGoing \; (s', v_1) \le s_2$. Then, it suffices to discuss the case where $v'' = v$ as it gives the least $s''$. In this case, we have $s'' = (A,D) = s$ and thus $s \le s''$.
When $s'$ is a \fstate, $\GET \; \filterOnGoing \;s'$ is also a \fstate. 
If $v''$ is a \fstate, \ie, $v'' = \GET \; \filterOnGoing \; s'$. Then, we have $s'' = (s' \setminus \SelOG{s'}) \lhd \SelOG{s'} = s'$, which implies $s \le s'$. 
If $v''$ is a \pstate, it again suffices to consider the case where $v'' = v$, which leads to $s'' = (A, D) = s$ and thus $s \le s''$.

\subsection{Well-Behavedness of \texorpdfstring{$\filterOnGoing$}{filterOnGoing} in \cref{sec:elaborated-domains-for-finer-grained-updates}}
\label{sec:well-behavedness-filterOnGoing}

Since $\filterOnGoing$ is still classically well-behaved when only \fstates are involved, we only need to examine the case where \spstates are involved.

First, we confirm \ref{law:aa-consistency}.
Suppose that $\PUT \; \filterOnGoing \; (s,v) = s'$. Since $\GET \; \filterOnGoing$ is monotone, it suffices to show $v \le \GET \; \filterOnGoing \; s'$. 
By case analysis on $v$, it is easy to see $v = \GET \; \filterOnGoing \; s'$ by definition of $\filterOnGoing$. 

Second, we confirm \ref{law:aa-acceptability}. 
Suppose that $\GET \; \filterOnGoing \; s = v$ and $\IDon{v'}{v}$. 
We will show $\PUT \; \filterOnGoing \; (s,v') = s'$ for some $s'$ with $\IDon{s'}{s}$. 
Consider the case where $s = t \in \FTables$. If $v' = v = \SelOG{t}$, then it is easy to see that $s' = (t \setminus \SelOG{t}) \lhd \SelOG{t} = t$, ensuring $s' = \IDon{t}{t}$. 
If $v' = (A, D) \in \IDonSet{\SelOG{t}}$, then it must be the case where $A \subseteq \SelOG{t} \in \FTablesOG$ and $D = \emptyset$, which implies $s' = (A,\emptyset)$. 
Since we have $A \subseteq \SelOG{t} \subseteq t$, we have $s' = (A, \emptyset) \in \IDonSet{t}$.
Consider the case where $s = (A,D)$. Then, it must be the case where $v = (\SelOG{A}, D)$. 
By definition, $v'$ that yields the greatest $s'$ is $v' = v$, and it suffices to consider the case where $v' = v$. 
In this case, $s' = (\SelOG{A}, D)$. Since we have $\SelOG{A} \subseteq A$, we have $s' = (\SelOG{A},D) \in \IDonSet{(A,D)}$.

Finally, we confirm \ref{law:aa-stability}. 
Suppose that $\PUT \; \filterOnGoing \; (s_0,v) = s$, and $\PUT \; \filterOnGoing \; (s', v'') = s''$ with $s \le s'$ and $v \le \IDon{v''}{\GET \; \filterOnGoing \; s'}$. We will show $s \le s''$. When $s$ is a \fstate, only \fstates are involved in the discussion. 
Hence, we consider the case where $s = v = (A,D)$ with $A \in \FTablesOG$. 
When $s' = (A', D')$, $\GET \; \filterOnGoing \; s'$ is also a \spstate. 
For \spstates $v_1, v_2$ with $v_1 \le v_2$, we have $\PUT \; \filterOnGoing \; (s', v_2) = s_2$ implies $\PUT \; \filterOnGoing \; (s', v_1) \le s_2$. Then, it suffices to discuss the case where $v'' = v$ as it gives the least $s''$. In this case, we have $s'' = (A,D) = s$ and thus $s \le s''$.
When $s'$ is a \fstate, $\GET \; \filterOnGoing \;s'$ is also a \fstate. 
If $v''$ is a \fstate, \ie, $v'' = \GET \; \filterOnGoing \; s'$. Then, we have $s'' = (s' \setminus \SelOG{s'}) \lhd \SelOG{s'} = s'$, which implies $s \le s'$. 
If $v''$ is a \pstate, it again suffices to consider the case where $v'' = v$, which leads to $s'' = (A, D) = s$ and thus $s \le s''$. 

\subsection{\cref{lemma:ran-respected-for-fine-enough-updates}}

Suppose that the following condition holds: for all $s' \in \Ran{s}{u_1} \cap \Ran{s}{u_2}$, there exists $u$ such that $u_1 \le_U u$, $u_2 \le_U u$ and $s' \in \Ran{s}{u}$.
Suppose also that $u_1 \oplus_U u_2 = u_{12}$ and $s' \in \Ran{s}{u_1} \cap \Ran{s}{u_2}$. 
Then, since $\oplus_U$ soundly implements the join, we have $u_{12} = u_1 \vee u_2$ and thus have $u_{12} \le_U u$ as the join is the least common upper bound of $u_1$ and $u_2$. 
By definition, $\Ran{s}{u}$ is antitone: $u \le_U u'$ implies $\Ran{s}{u'} \subseteq \Ran{s}{u}$ for any $u$ and $u'$. Thus, by $s' \in \Ran{s}{u}$ and $u_{12} \le_U u$, we also have $s' \in \Ran{s}{u_{12}}$, which proves that $\oplus_U$ respects $\Ran{s}{-}$ for any $s$.

\subsection{\cref{lemma:pijs-holds-for-associatve-join}}

Suppose that $u_1 \oplus_U u_2$ is defined for any $u_1$ and $u_2$ with $u_1 \le u_2$, and that $u_1 \oplus_U (u_2 \oplus_U u_3) = u$ implies $(u_1 \oplus_U u_2) \oplus_U u_3 = u$ and vice versa for any $u_1, u_2, u_3, u \in U$.
Notice that $u_1 \oplus_U u_2 = u$ for $u_1 \le u_2$ implies $u = u_2$. 
Let $u_1$ and $u_2$ be updates with $u_1 \le_U u$ and $u_2 \le_U u$ for some $u$. 
Then, we prove that $u_1 \oplus_U u_2$ must be defined. We use the fact that $u_1 \oplus_U u = u$ and $u_2 \oplus_U u = u$ as we assumed $(\le) \subseteq \dom(\oplus_U)$. 
We have:
\begin{align*}
 u_1 \oplus_U u = u 
 & \Leftrightarrow \text{\{ $u_2 \oplus_U u = u$ by assumption \}} \\
 & \phantom {{}\Leftrightarrow{}} u_1 \oplus_U (u_2 \oplus_U u) = u \\
 & \Leftrightarrow \text{\{ associativity \}} \\
 & \phantom {{}\Leftrightarrow{}} (u_1 \oplus_U u_2) \oplus_U u = u 
\end{align*}
Since $(u_1 \oplus_U u_2) \oplus_U u$ is defined, $u_1 \oplus_U u_2$ must also be defined. 

Note that we also have $u_1 \oplus_U u_2 \le u$.

\section{A Lawful Lens Whose $\PUT$ is not Monotone in the First Argument}
\label{sec:non-monotone-in-first-arg}

\newcommand{\PutNonmonoFst}{\ensuremath{\ell_\mathrm{putNonmono1}}\xspace}

Consider the following lens $\PutNonmonoFst \in \MALens{\con{Bool}_\GO}{\UnitType_\GO}$.
\[\bb
\GET \; \PutNonmonoFst \; s = \begin{cases} 
\GO & \text{if}~s = \GO \\
(\,) & \text{otherwise} 
\end{cases}
\\
\PUT \; \PutNonmonoFst \; (s,v) = \begin{cases}
\GO &\text{if}~v = \GO \\
s   &\text{if}~v \ne \GO \wedge s \ne \GO \\
\TT &\text{if}~v \ne \GO \wedge s = \GO 
\end{cases} 
\ee\]
This lens is not monotone in the first argument because we have $\PUT \; \PutNonmonoFst \; (\GO, (\,)) = \TT$ while $\PUT \; \PutNonmonoFst \; (\FF, (\,)) = \FF$, where $\TT \not\le \FF$. Notice that this $\GET$ is monotone and $\PUT$ is also monotone in the second argument. 

Let us confirm that this lens is well-behaved. First, we confirm \ref{law:aa-acceptability}. 
Since its $\PUT$ is monotone in the second argument, it suffices to check $\IDon{\PUT \; \PutNonmonoFst \allowbreak\; (s, \GET \; \PutNonmonoFst \; s)}{s}$. When $s = \GO$, the left-hand side is also $\GO$ by definition, satisfying $\IDon{\GO}{\GO}$. When $s \ne \GO$, the left-hand side is $s$ by definition, satisfying $\IDon{s}{s}$. 
Next, we confirm \ref{law:aa-consistency}. Since its $\GET$ is total and monotone, it suffices to check $v \le \GET \; \PutNonmonoFst \; (\PUT \; \PutNonmonoFst \; (s,v))$. Since this is trivial if $v = \GO$, we focus only on the case where $v = (\,)$. In this case, by definition, we have $s' = \PUT \; \PutNonmonoFst \; (s,v) \ne \GO$ by definition, and thus $\GET \; \PutNonmonoFst \; s' = (\,)$. 
Finally, we confirm \ref{law:aa-stability}. Suppose that $\PUT \; \PutNonmonoFst \; (s_0, v) = s$ and $\PUT \; \PutNonmonoFst \; (s', v'') = s''$ with $s \le s'$ and $v \le \IDon{v''}{ \GET \; \PutNonmonoFst \; s'}$. We will show $s \le s''$. Since its $\PUT$ is monotone in the second argument, it suffices to consider the case of $v'' = v$. When $v = \GO$, we have $s = s'' = \GO$ by definition. 
Suppose $v = (\,)$. When $s_0 = \GO$, we have $s = s' = s'' = \TT$ by definition. 
When $s_0 \ne \GO$, we have $s = s' = s'' = s_0$ by definition.

\section{A Well-Behaved Lens that Does not Satisfy \textsc{WPutGet}}
\label{sec:appendix-lens-that-does-not-satisfy-WPutGet}

\newcommand{\ConstZns}{\var{const0}_\mathrm{ns}}
\newcommand{\ConstUnit}{\var{constUnit}_\mathrm{ns}}

Let $\UnitType_\Omega$ be the free pointed poset obtained from $\UnitType = \{(\,)\}$. Consider the following lens $\ConstUnit \in \MALens{\UnitType_\GO}{\UnitType_\GO}$.
\[
\GET \; \ConstUnit \; \dontcare = (\,) 
\qquad 
\PUT \; \ConstUnit \; (s,v) = \begin{cases}
 s    & \text{if}~v = (\,) \\
 \GO  & \text{if}~v = \GO 
\end{cases}
\]
The lens is well-behaved: it satisfies \ref{law:aa-acceptability}, \ref{law:aa-consistency} and \ref{law:aa-stability}. We only show the \ref{law:aa-stability} as the other two are more straightforward. 
Suppose that $\PUT \; \ConstUnit \; (s_0, v) = s$, $s \le s'$, $v \le \IDon{v''}{\GET \; \ConstUnit \; s'}$, and $\PUT \; \ConstUnit \; (s',v'') = s''$. Notice that $\GET \; \ConstUnit \; s' = (\,)$ regardless of $s'$. 
We will show $s \le s''$ by case analysis on $v$.
Consider the case $v = (\,)$. In this case, we have $s = s_0$. Also, $v = (\,) \le \IDon{v''}{(\,)}$ implies $v'' = (\,)$. Then, we have $s'' = \PUT \; \ConstUnit \; (s,v'') = s'$ by definition. Hence, we have $s \le s'' = s'$ by the assumption $s \le s'$. 
Consider the case $v = \GO$. In this case, we have $s = \GO$. Since $\GO$ is the least element in $\UnitType_\GO$, we have $s = \GO \le s''$ regardless of $s''$.

In contrast, $\ConstUnit$ does not satisfy \ref{law:wput-get}. We have
$\PUT \; \ConstUnit \allowbreak\; ((\,), \Omega) = \Omega$, but we also have
$\PUT \; \ConstUnit \; ((\,), \GET \; \ConstUnit \; \Omega) =
 \PUT \; \ConstUnit \; ((\,), (\,)) = (\,) \ne \Omega$.

 \fi

\end{document}